\documentclass{pasa}%

\usepackage[authoryear]{natbib}
\bibpunct{(}{)}{;}{a}{}{,}
\usepackage{graphicx}
\usepackage{amsmath}
\usepackage{gensymb}
\usepackage{aas_macros}
\usepackage{url}
\usepackage{color}

\newcommand{\red}[1] {#1}
\newcommand{\rednew}[1] {#1}

\title[BIGHORNS]{BIGHORNS - Broadband Instrument for Global HydrOgen ReioNisation Signal}
\author[Marcin Sokolowski et al.]
{Marcin Sokolowski$^{1,2}$\thanks{marcin.sokolowski@curtin.edu.au},       
 Steven E. Tremblay$^{1,2}$,           
 Randall B. Wayth$^{1,2}$,          
 Steven J. Tingay$^{1,2}$,          
 Nathan Clarke$^1$,             
 Paul Roberts$^3$,              
 Mark Waterson$^{1,4}$,             
 Ronald D. Ekers$^3$,           
 Peter Hall$^1$,                
 Morgan Lewis$^5$,
 Mehran Mossammaparast$^{1,2}$, 
 Shantanu Padhi$^1$,            
 Franz Schlagenhaufer$^1$,      
 Adrian Sutinjo$^1$,            
 \and Jonathan Tickner$^{1}$ \\ 
\\
\affil{$^1$International Centre for Radio Astronomy Research, Curtin University, GPO Box U1987, Perth, WA 6845, Australia} 
\affil{$^2$ARC Centre of Excellence for All-sky Astrophysics (CAASTRO), Redfern, NSW, Australia}                      
\affil{$^3$CSIRO Astronomy and Space Science, PO Box 76, Epping, NSW 1710, Australia}                                 
\affil{$^4$SKA Organisation, Jodrell Bank Observatory, Lower Withington, Macclesfield, SK11 9DL, United Kingdom}
\affil{$^5$International Centre for Radio Astronomy Research, University of Western Australia, 35 Stirling Highway, Perth, WA 6009, Australia}
}%
\jid{PASA}
\doi{10.1017/pas.\the\year.xxx}
\jyear{\the\year}

\begin{document}%
\begin{abstract}
\red{The redshifted 21cm line of neutral hydrogen (\textsc{Hi}), potentially observable at low radio frequencies ($\sim50-200$\,MHz), should be a powerful probe
of the physical conditions of the inter-galactic medium during Cosmic Dawn and the Epoch of Reionisation (EoR).}
The sky-averaged \textsc{Hi} signal is expected to be extremely weak ($\sim$100\,mK) in comparison to the foreground of up to $10^4$\,K at the lowest frequencies of interest.
The detection of such a weak signal requires an extremely stable, well characterised system and a good understanding of the foregrounds. \red{Development of a nearly perfectly ($\sim$mK accuracy) calibrated  total power radiometer system is essential for this type of experiment.} 
We present the BIGHORNS (Broadband Instrument for Global HydrOgen ReioNisation Signal) experiment which was designed and built to detect the sky-averaged \textsc{Hi} signal from the EoR at low radio frequencies.
The BIGHORNS system is a mobile total power radiometer, which can be deployed in any remote location in order to collect radio-interference (RFI) free data.
\red{The system was deployed in remote, radio quiet locations in Western Australia and low RFI sky data have been collected.
We present a description of the system, its characteristics, details of data analysis and calibration. We have identified multiple challenges to achieving the required measurement precision, which triggered two major improvements for the future system.}
\end{abstract}
\maketitle%
\section{INTRODUCTION }
\begin{figure}
  \begin{center}
	 \includegraphics[width=3in]{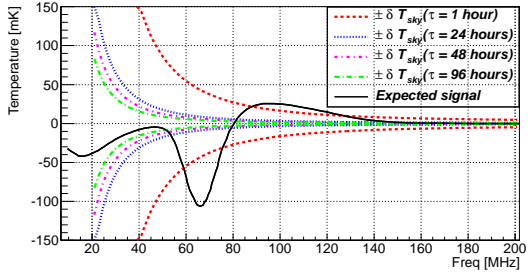}
    \caption{\red{The expected global EoR signal \citep{2010PhRvD..82b3006P} with expected sky noise (assuming ideal noiseless receiver) as a function of frequency calculated according to formula $\delta T_{sky} = T_{sky} / \sqrt{B \tau}$, for frequency resolution B=1\,MHz, integration times $\tau=$ 1\,h, 24\,h, 48\,h and 96\,h, and the expected sky temperature estimated according to formula $T_{sky} = $180\,K $(180/\nu)^{2.6}$, where $\nu$ is frequency in MHz.}}
    \label{fig_expected_global_eor}
  \end{center}
\end{figure}

\begin{figure}
  \begin{center}
	 \includegraphics[width=3in]{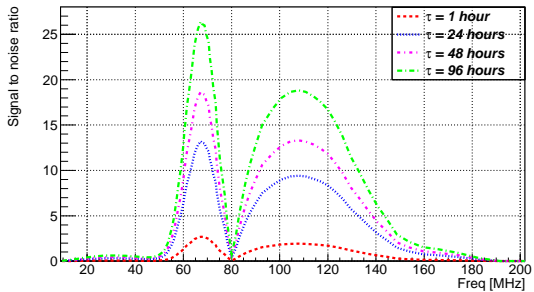}
    \caption{\red{Expected signal to noise ratio derived from Figure~\ref{fig_expected_global_eor} for the same integration times. Although, the absorption trough at $\sim60-80$\,MHz looks like the easiest spectral feature to detect, the actual signal to noise ratio is not significantly lower at the emission ``bump''. Therefore, it might be easier to detect the emission peak at higher frequencies, which are less affected by ionospheric effects and thus better understood. }}
    \label{fig_snr}
  \end{center}
\end{figure}

Determining what happened in the early Universe before the Epoch of Reionisation (EoR) is one of the high priority goals of modern cosmology.
A wealth of information on the physics of reionisation is encoded in the redshifted 21\,cm signal from neutral hydrogen (\textsc{Hi}) during these times (see e.g. \citet{2012RPPh...75h6901P} for a recent review). 
\red{The frequency of the 21\,cm line (1420\,MHz) redshifts into the low radio frequency range of $200-50$\,MHz, for $6<z<30$.}
The challenge of detecting the redshifted 21\,cm signal is being met on several fronts. Interferometer arrays [MWA: \citealt{2013PASA...30....7T}; LOFAR: \citealt{2013A&A...556A...2V}; LWA: \citealt{2012JAI.....150004T}; PAPER: \citealt{2010AJ....139.1468P}] aim for a statistical detection of the \textsc{Hi} power spectrum and the Square Kilometre Array (SKA: \citealt{2013ExA....36..235M}) plans to perform full tomography of the EoR \citep{AASKA14}.
On the other hand, there are also attempts to identify the EoR signature in the integrated spectrum of the whole sky at low radio frequencies -- the ``global'' EoR signal.
The physics of the cosmological 21cm line and expectations for the global signature have been described in detail by number of authors \red{\citep{1999A&A...345..380S, 2006PhR...433..181F, 2010ARA&A..48..127M, 2008PhRvD..78j3511P, 2010PhRvD..82b3006P, 2012RPPh...75h6901P}}. The main expected features of the global signal are a trough of absorption against the CMB around 70\,MHz and a peak of emission against the CMB around 100\,MHz.
Depending on the cosmological model, the depth of the trough is $\sim100$\,mK and emission peak $\sim30$\,mK (Fig.~\ref{fig_expected_global_eor}) against a typical sky temperatures of around 2000\,K at $\sim$70~MHz, decreasing to $\sim$300\,K at $\sim$150\,MHz. 
\red{In the case of a sky dominated receiver (no noise contribution from the receiver itself), the expected signal to noise ratio after 24 hours of integration exceeds (at the absorption trough) or is close to (at the emission peak) 10 at frequencies corresponding to the absorption trough and emission peak respectively (Fig.~\ref{fig_snr}).}
\red{A guide of how to design an experiment to enable data analysis capable of distinguishing the global 21\,cm signal from foreground contaminants was given by \citet{2013PhRvD..87d3002L}}.

\begin{figure*}
  \begin{center}
	 \includegraphics[width=6in]{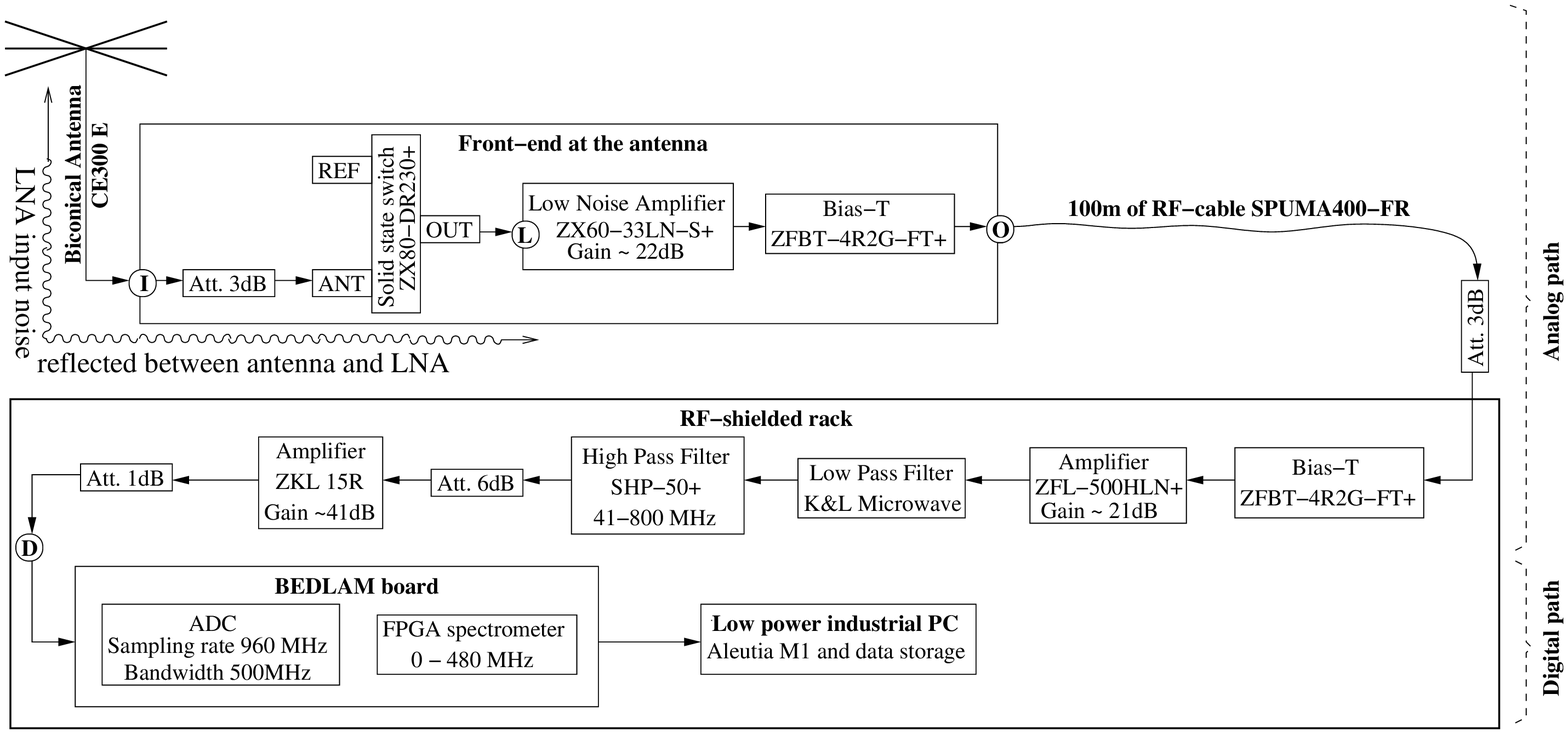}
    \caption{The BIGHORNS RF signal path as it was deployed in the field in 2012-2014. \red{The antenna and the front-end (with a 3~dB attenuator in between) were separated by 100~m of RF-cable from the back-end electronics. The back-end comprised of a second gain stage (amplifiers separated by RF-filters), digitiser and spectrometer implemented on the BEDLAM board and a low power industrial PC for data acquisition. The three reference points marked in the image are: \textbf{I} - input of the front-end, \textbf{L} - input of the first LNA, \textbf{O} - output of the front-end and \textbf{D} - beginning of the digital part of the system. Unless stated otherwise, we will further refer to the receiver (receiver noise temperature etc.) as the entire analogue signal chain from the reference point \textbf{I} down to \textbf{D}, whilst to LNA as the first amplifier in the signal path (ZX60-33LN-S+).}}  
    \label{fig_ebo_setup}
  \end{center}
\end{figure*}





\red{In principle, measurement with a precision of the order of 10\,mK can be achieved in $\sim$24 hours of night time observations when the Galactic Centre is below the horizon (Fig.~\ref{fig_expected_global_eor}).
In practice, depending on the time of the year, this translates to at least a few days of observations to be able to average only RFI-excised data obtained in optimal conditions.}
\red{Several groups [EDGES: \citealt{2010Natur.468..796B}; CoRE: \citealt{2009PhDT.......301C}; SARAS: \citealt{2013ExA....36..319P}, SCI-HI: \citealt{2014ApJ...782L...9V}, LEDA: \citealt{2014era..conf10301G}, DARE: \citealt{2012AdSpR..49..433B}] are planning or operating experiments to detect the global EoR signal.}
\red{The EDGES experiment has already placed a lower limit ($\Delta z > 0.06$) for the duration of reionisation \citep{2010Natur.468..796B}, whilst limits on the kinetic Sunyaev-Zel'dovich (kSZ) effect from the South Pole Telescope yielded the upper limit ($\Delta z < 7.9$) \citep{2012ApJ...756...65Z}}.

Interest in the global EoR signal has motivated work to explore how well the data from these experiments can constrain reionisation scenarios \citep{2012MNRAS.419.1070H, 2012MNRAS.424.2551M, 2013PhRvD..87d3002L, 2013ApJ...777..118M} and how instrumental effects affect the results \citep{2014arXiv1404.0887B}.

\red{A nearly perfect understanding of the signal path, response and calibration of the instrument is essential for the global EoR experiment and for any sophisticated data analysis which has to be performed to identify the global EoR signature.}
Multiple instrumental effects that must be controlled or understood to high precision have already been discussed by \citet*{2012RaSc...47.0K06R}. Perhaps the most important of these is the stability of the system's response over several days of integration time.
Integrating down to the desired precision requires a very stable radiometer system deployed at a very radio quiet site to avoid human-made radio frequency interference (RFI).
In order to be able to identify the global EoR signal in the radio spectrum the system's response must be smooth and understood to a mK level, likewise the galactic foreground radio emission.

\red{Recent studies have shown that ionospheric absorption and refraction can significantly affect the frequency structure of the cosmic signal \citep{2014MNRAS.437.1056V,2014arXiv1409.0513D}.
Furthermore, the intra-night variability of the electron content in the ionosphere (even under ``quiet'' night time conditions) may impede detection of the global EoR spectral signature from the ground \citep{2014arXiv1409.0513D}.}


\red{This paper summarises the current status of the BIGHORNS experiment and discusses several topics (instrument, RFI, data processing, calibration and modelling) important to every global EoR experiment. In Section~\ref{sec_instrument_desc} we present the BIGHORNS system, its RF-signal path and characteristics.
In Section~\ref{sec_data_processing} we present the data collected in several remote radio quiet locations of Western Australia and describe processing, quality assessment, calibration and modelling of the these data.
In Section~\ref{sec_system_limitations} we present studies of the system stability and its current limitations which triggered modifications to the signal path, described in Section~\ref{sec_future_improvements}. 
In Section~\ref{sec_future_improvements} we also present the conical log spiral antenna, which was developed specifically for the BIGHORNS experiment and will be used in future deployments.
}

\section{Instrument Description}
\label{sec_instrument_desc}
\begin{figure}
  \begin{center}
    \leavevmode
	 \includegraphics[width=3in]{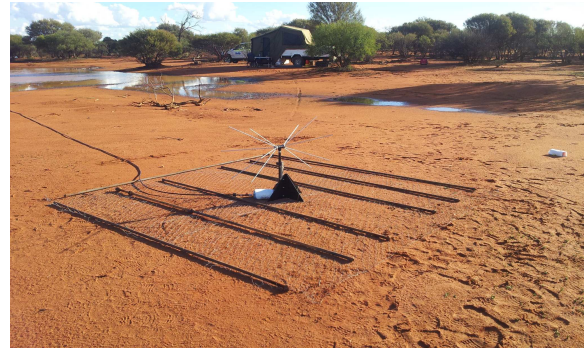}
    \caption{\red{The BIGHORNS antenna and the front-end box (in the foreground) separated by 100\,m of RF-cable from the camper trailer housing an RF-shielded rack with the back-end electronics (in the background) as deployed at the Wondinong Station in April 2014.}}
    \label{fig_wondinong_photo}
  \end{center}
\end{figure}

The BIGHORNS system is a total power radiometer. 
\red{The main driving forces motivating the design of the BIGHORNS prototype were: simplicity of the signal path; portability; the possibility of deployment in remote locations without access to electricity; and low power consumption.}
The schematic of the system is presented in Figure~\ref{fig_ebo_setup}.
The system consists of a broadband off-the-shelf biconical antenna (CE300E by Compliance Engineering Pty Ltd) mounted 52\,cm above a $3\times3$\,m wire mesh (pitch size $\approx$55\,mm) ground screen (Fig.~\ref{fig_wondinong_photo}), 
which receives radio waves from a large fraction of the sky. During all deployments the antenna was oriented in the East-West direction \red{(Fig.~\ref{fig_wondinong_photo})}.


The signal intercepted by the antenna is amplified in the analogue chain, which presently consists mainly of off-the-shelf components.
\red{The total gain of the system ($\sim 68-75$\,dB) was set in order to maximise occupancy of the 8 bit dynamic range of the digital system with some headroom.
The signal path starts with the front-end box. A 3~dB attenuator is inserted between the antenna and the front-end box in order to improve the impedance match and suppress reflections of the LNA input noise between the two, as described by \citet*{2012RaSc...47.0K06R}.
In the case of the prototype system with the biconical antenna, which is well matched only over a relatively narrow frequency band, the attenuator provided a better match over a wider frequency band.
However, it is undesired because it increases the noise temperature of the receiver (reduces the signal to noise ratio) and is planned to be removed (improvements in the presented system will be discussed in Section~\ref{sec_future_improvements}).}
The front-end box is powered via the Bias-T (Mini-Circuits part number \footnote{The codes in the brackets refer to Mini-Circuit part numbers, unless stated otherwise} ZFTB-4R2G-FT+) and houses the first low noise amplifier (\red{ZX60-33LN-S+ with gain $\sim$22\,dB and noise figure (NF)~$\sim$1.1\,dB}) and a two positional solid-state switch (ZX80-DR230+\red{, loss $\sim$0.75\,dB}).
The switch is controlled by an NE555 chip-based circuit to switch between the antenna and a reference source in order to calibrate the RF-signal in temperature units and also to calibrate gain variations. 
The reference source is a 50\,$\Omega$ terminator at ambient temperature. The temperature is measured with a logger deployed in an RF-shielded box similar to the one housing the front-end components.
The switching cadence can be set to two predefined configurations, $\approx$15.5\,s on antenna and 5s on the reference, or $\approx$7\,s on antenna and 5\,s on reference.

\begin{figure}
  \begin{center}
    \leavevmode
	 \includegraphics[width=3in]{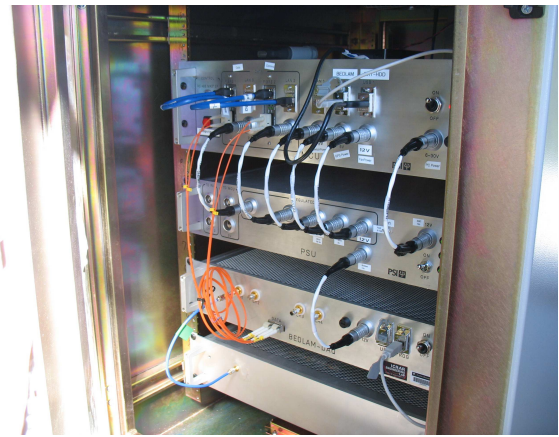}
    \caption{The RF-shielded rack housing BIGHORNS back-end components (from the bottom to top : box with a second gain stage and RF-filters; BEDLAM spectrometer; power unit; and PC computer unit).}
    \label{fig_rack_ebo}
  \end{center}
\end{figure}

From the Bias-T the RF-signal is carried over 100\,m of SPUMA-400-FR cable \red{(loss $\sim$2-7\,dB in the frequency range 20-300~MHz)} to the RF-shielded rack (Fig.~\ref{fig_rack_ebo}).
A 3\,dB attenuator is inserted at the end of the 100\,m cable in order to suppress reflected signals. Inside the rack the signal goes to a box where a second stage of gain and RF-filters are installed (Fig.~\ref{fig_ebo_setup}).
\red{The second stage of amplification consists of amplifiers ZFL-500HLN+ (gain~$\sim$21\,dB, NF$\sim$3.9\,dB) and ZKL-1R5 (gain~$\sim$41\,dB, NF~$\sim$2.8\,dB).}
In order to band limit the signal and suppress high power signals below $\sim20$\,MHz, a high pass (41-800\,MHz) filter (SHP-50+) and a low pass (0-300\,MHz) K\&L Microwave filter (Microwave part number 7L120-306/T900-0/0) are used.
\red{The highest frequencies of interest is $\sim$250\,MHz, but having a slightly higher cut-off provides a potentially useful diagnostic band and ensures smoothness of the system's response.}
The signal is then passed to the BEDLAM board \citep{2013ExA....36..155B}, which was originally developed for the LUNASKA experiment \citep{2011MNRAS.410..885J} and consists of analogue-to-digital converters (ADCs) and FPGAs configured as spectrometers.
\red{
The BEDLAM board is a good option for a low-power-consumption spectrometer ($< 60$\,W). The bandwidth of the ADC is 500 MHz and the signal is sampled at 960 MSamples/sec and digitised at 8 bits precision. The resulting sampled frequency range is more than strictly required given the 300 MHz lowpass filter, however the empty spectrum between approximately 350 and 450 MHz in our data provides a useful diagnostic window for variations in the signal noise floor due to analogue or digital effects.
}
The spectrometer implementation uses a four tap polyphase filterbank (PFB) and channelises the data into 4096 channels with spectral resolution $\approx$117.2\,kHz.
The power spectrum is formed and initially accumulated in the FPGA and the number of accumulations, hence time resolution, is configurable.
Typically we have collected data with $\sim$270\,ms and 50\,ms time resolutions.
Finally, the accumulated spectra are transmitted over an ethernet interface to a low power industrial PC computer (Aleutia M1) which saves the spectra on a hard drive in a FITS format \citep{2010A&A...524A..42P}.
The data rate at 50\,ms time resolution is about 28.5\,GB per 24\,hours. The off-line data processing is described in Section~\ref{sec_data_processing}.

\begin{figure}
  \begin{center}
    \leavevmode
	 \includegraphics[width=3in]{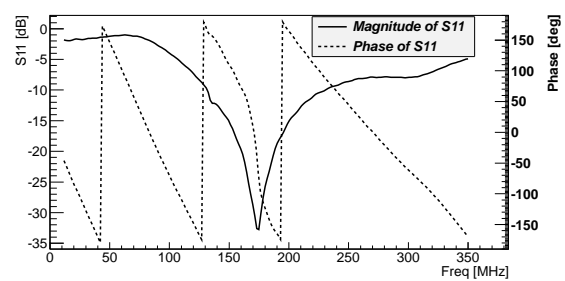}
    \caption{The magnitude and phase of the reflection coefficient of the antenna-over-groundscreen as deployed at Wondinong Station \red{measured at the reference point \textbf{I} (Fig.~\ref{fig_ebo_setup})} .} 
    \label{fig_antenna_s11}
  \end{center}
\end{figure}

\subsection{Reflection coefficients of the components}
\label{subsec_comp_char}


\red{The scattering parameters (S-parameters) characterise the response of RF-components to RF-signals as a function of frequency. The reflection coefficient (S11) is a complex ratio of the reflected to incident voltage. 
Therefore, its squared magnitude ($|S11|^2$) represents the fraction of power reflected and is in the range [0,1], which translates into $(-\infty,0]$ in dB scale.
The transmission coefficient (S21) is a complex ratio of voltages at the output and input of a device. Thus, its squared magnitude is a measure of gain ($>0$\,dB) or attenuation ($<0$\,dB).
$|S11|^2$ and $|S21|^2$  are typically expressed in dB, and we will also typically refer to reflection coefficient as magnitude expressed in dB scale.
}
Characteristics (such as the S-parameters or the noise figure) of the components were measured with a Vector Network Analyser (VNA) whenever it was necessary.
During field trips, the antenna's reflection coefficient was measured with a portable Rohde-Schwarz ZVL VNA,
whilst for more precise measurements (such as noise temperature), which could be done in the laboratory, a Agilent PNA-X was used.
\red{The reflection coefficient of the antenna deployed at the Wondinong Station in April 2014 (see Section~\ref{subsec_data_collection}) was measured at the reference point \textbf{I} (Fig.~\ref{fig_ebo_setup}). Its magnitude and phase are shown in Figure~\ref{fig_antenna_s11}.
The biconical antenna is very well matched to the front-end \mbox{($|\Gamma_{al}|^2<$~0.01)} only in the frequency band $\sim$172-184\,MHz and reasonably well matched \mbox{($|\Gamma_{al}|^2<$~0.1)} across the rest of the $\sim130-230$\,MHz band.
In the prototype stage of the experiment the antenna was required to be portable. Therefore, its characteristics were a compromise between what was available on the market and the predictions for the global EoR signal at the time of purchase.
The requirements for the reflection coefficient of the antenna will be discussed in Section~\ref{subsec_antenna_mismatch}.}

\begin{figure}
  \begin{center}
    \leavevmode

	 \includegraphics[width=3in]{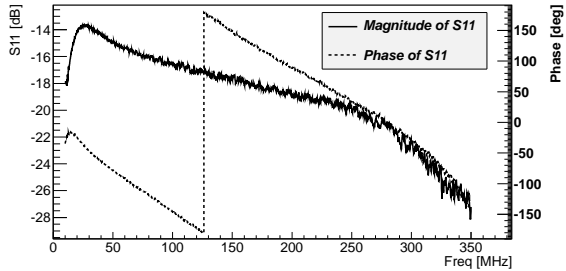}
    \caption{The magnitude and phase of the reflection coefficient of the front-end (with a 3\,dB attenuator connected to its input) \red{measured at the reference point \textbf{I} (Fig.~\ref{fig_ebo_setup})}  .}
    \label{fig_frontend_s11}
  \end{center}
\end{figure}

\red{The reflection coefficient of the front-end was measured at the reference point \textbf{I} (Fig.~\ref{fig_ebo_setup}). Its magnitude and phase are shown in Figure~\ref{fig_frontend_s11}}.

\begin{figure}
  \begin{center}
    \leavevmode
	 \includegraphics[width=3in]{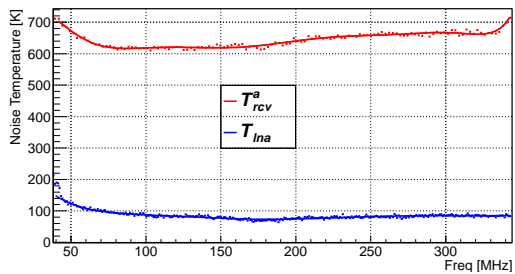}
    \caption{Measured noise temperatures of the low noise amplifier ZX60-33LN-S+ \red{(blue curve)} and of the entire signal path with a 3\,dB attenuator at the antenna input of the front-end with a 9$^{th}$ order polynomial fit superimposed \red{(red curve)}.}
    \label{fig_tlna}
  \end{center}
\end{figure}


\begin{figure}
  \begin{center}
    \leavevmode
    \includegraphics[width=3in]{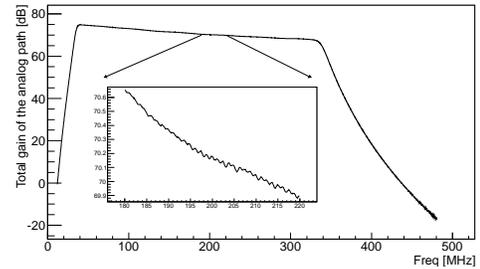}
    \caption{The total gain of the analogue path measured with the Rohde-Schwarz ZVL Vector Network Analyser. A quasi-periodic ripple of expected periodicity $\approx$1.275\,MHz due to a standing wave in the 100\,m cable is shown in a frequency range 180-220\,MHz.}
    \label{fig_total_gain}
  \end{center}
\end{figure}

\subsection{Receiver gain and noise temperature}
\label{subsec_noise_temp}

\red{The noise temperature (Fig.~\ref{fig_tlna}) and the gain of the whole system (Fig.~\ref{fig_total_gain}) were measured between the reference points \textbf{I} and \textbf{D} (Fig.~\ref{fig_ebo_setup}).} The noise temperature of the system was measured in the laboratory with the PNA-X prior to the astronomical observations. 
The noise temperature of the terminated, standalone low noise amplifier is $T_{lna} \approx$ 80\,K, the noise temperature of the entire signal path without a 3\,dB attenuator is $\approx$ 170\,K, whilst the noise temperature of the entire signal path with a 3\,dB attenuator is $T^{a}_{rcv} \approx$ 600\,K. 
\red{The receiver noise temperatures (without and with the attenuator) agree with the expected values calculated according to the Friis formula ($\sim$160~K and $\sim$620~K respectively) for the system presented in Figure~\ref{fig_ebo_setup}.}
The noise temperatures were parameterised by a 9$^{th}$ order polynomial in order to be used in the calibration process (Fig.~\ref{fig_tlna}). 
\red{Based on the reduced $\chi^2$ test, the 9$^{th}$ order polynomial was sufficient to describe the noise temperature data (measurement error $\sim$5.4~K) in $50-340$\,MHz band. Nevertheless, data analysis may be performed in the narrower frequency band where lower order polynomials should be sufficient. In a single (cold state) calibration schema, without the possibility of continuous determination of the noise temperature from hot/cold reference sources, higher accuracy noise temperature measurements are also required.}

\subsection{System portability}

In order to be able to deploy the system in any remote location without access to electricity it was designed to be portable. In the field it is deployed in a modified off-road camper trailer.
The whole system consumes about 120\,W of power.
In order to make the system fully mobile, the power system consisting of a 24\,V, 130\,Ah battery, 24\,V~/~23\,A smart charger, and a power generator was assembled.
The capacity of the battery is sufficient to continuously run the system overnight.
In the field the system was running on the battery during the night and the battery was charged for a few hours during the day from a power generator (Honda 20EU).

\subsection{Electromagnetic compatibility testing}
The self-generated electromagnetic interference (EMI) from the system was measured in an anechoic chamber at Curtin University.
\red{The emissions from the systems were compared against the military standard MIL-STD461F \citeyearpar{MILSTD} and commercial standard CISPR 22 Class B
\citeyearpar{CISPR_STD}, which define a well established procedure of testing electronic equipment. Moreover, the EMI requirements of radio-quiet sites, such as the Murchison Radio-astronomy Observatory (MRO) in Western Australia,
are defined with respect to the military standard (i.e. 20\,dB below), and the system has to satisfy these requirements if it is ever intended to be deployed there. }
During the first two tests, EMI from the whole system was measured. 
In the first case the system was running on the battery with the charger turned off, which was a typical night time operation mode in the field. 
In this configuration the system satisfies the military standard MIL-STD461F \citeyearpar{MILSTD}. However, a few narrow band emissions in the $\sim150-160$\,MHz band and above 230\,MHz, and wide band emissions at $\sim$150\,MHz and above 190\,MHz were identified. 
Three narrow band emissions above 220\,MHz were close the military standard limit. Therefore, using these data and a model of radiowave propagation over the surface of the Earth (ITU-R P526-12 \citeyear{ITUR_MODEL}) we estimated the resulting power that could be picked up by the antenna as deployed at Wondinong Station (see Section~\ref{subsec_data_collection}).
In this configuration the antenna was placed at a distance of approximately 50\,m from the trailer housing the RF-shielded rack and the power system.
In the procedure we used sensitivity data (gain and antenna factor) measured by the manufacturer at 3\,m distance and provided with the antenna.
According to the antenna pattern obtained from the FEKO 6.3 electromagnetic simulation software, we estimated that the gain of the antenna above the ground screen in the direction of the rack (as deployed at Wondinong Station) was $\approx20-25$\,dB smaller than the gain provided in the antenna's datasheet.
Finally, we expressed the detected power in temperature units for 117.2\,kHz frequency bins.
Except for a few narrow band spikes reaching $\approx80-100$\,mK, which should be excluded from data analysis, the EMI power calculated at the antenna does not exceed $\sim$10\,mK in the frequency band 50-230\,MHz.
Thus, we conclude that our measurements at Wondinong Station were not significantly affected by self-generated RFI from the rack during the night time observations.

During the second test the charger was plugged into 240\,V socket and continuously charged the battery. Although the charger is the noisiest component of the system, it satisfies the commercial standard CISPR 22 Class B \citeyearpar{CISPR_STD}.
Such a configuration was not used in the field, where the battery was charged by the Honda generator (likely the noisiest part of the system).
However, this test allowed us to assess the EMI when the system is run off the 240\,V socket inside a shielded building, which may be the case in the future.

During the last test only the antenna, the front-end, and temperature probe in the metal box were inside the chamber and the rest of the system was outside.
No emission could be detected in this configuration down to the sensitivity limit of the EMI receiver of
the order of -10 -- 0\,dB$\mu$V/m in the frequency band $20-300$\,MHz. It is very likely that these simple components are much quieter than the sensitivity limit.
The front-end box is so close to the antenna that the upper limit translated into temperature units in the 117.2\,kHz frequency bin is of the order of Kelvins, which is insignificant in comparison to the requirements of the EoR experiment.
Nevertheless, the measurement gave us confidence that there is no significant undesired emission from the ``near antenna components''.


\section{Data collection, processing and calibration}
\label{sec_data_processing}

\subsection{Data collection}
\label{subsec_data_collection}

\begin{figure}
  \begin{center}
    \leavevmode
    \includegraphics[width=3in]{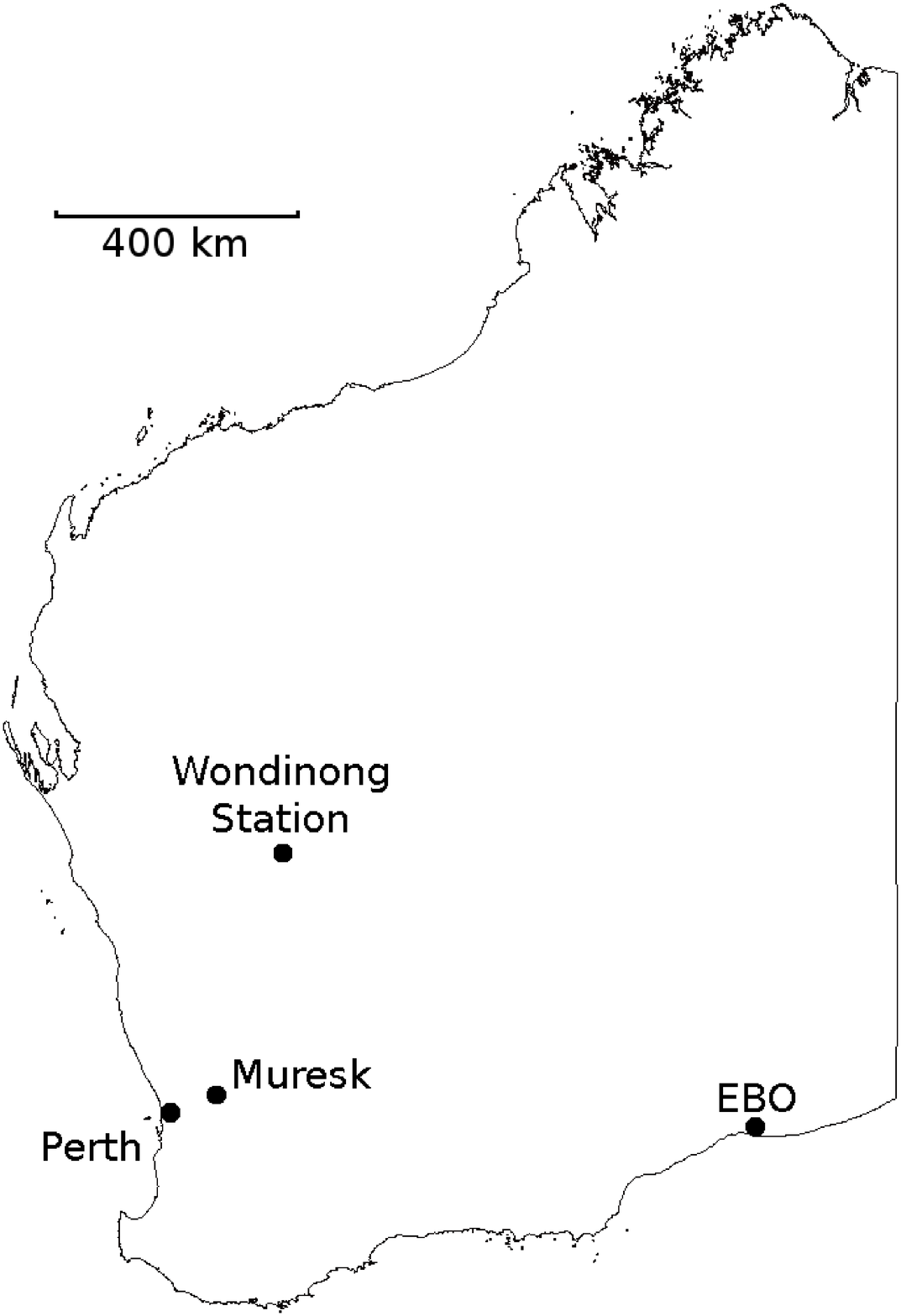}
    \caption{Outline map of Western Australia with Perth and the three deployment sites (Tab.~\ref{tab_sites}) marked.}
    \label{fig_wa_map}
  \end{center}
\end{figure}

Several datasets have been collected from three different locations in Western Australia (Tab.~\ref{tab_sites}). A number of substantial test datasets were collected in 2012 at Muresk (Fig.~\ref{fig_wa_map}). 
\red{While Muresk is an adequate test site (relatively close to Perth), the data quality is not good enough for global EoR studies. This is mainly due to RFI from relatively close transmitters in the FM (87-108\,MHz) and digital radio and TV bands (174-230\,MHz), which can saturate the receiver or cause significant distortions in the sky power spectra.}

\begin{table*}
\caption{The summary of the locations and major datasets collected with the BIGHORNS system in 2012-2014 period.}
\begin{center}
\begin{tabular}{@{}cccc@{}}
\hline\hline
Site & Geographic Coordinates & Duration & \begin{tabular}{@{}c@{}} Approximate amount \\ of collected data \end{tabular} \\
\hline%
 Muresk                      & 116$\degree$41'09''E , 31$\degree$44'46''S & \begin{tabular}{@{}c@{}}2012-10-04 - 2012-11-05 \\ 2014-02-08 - 2014-02-18\end{tabular} & \begin{tabular}{@{}c@{}} 32 days \\ 10 days\end{tabular}\\
 Eyre Bird Observatory (EBO) & 126$\degree$17'52''E , 32$\degree$15'08''S & 2013-12-07 - 2013-12-17 & 7 days \\
 Wondinong Station           & 118$\degree$26'24''E , 27$\degree$51'10''S & 2014-04-04 - 2014-04-11 & 5 days\\
\hline\hline
\end{tabular}
\end{center}
\label{tab_sites}
\end{table*}


\red{The quality of the data is another critical requirement for a measurement of the global EoR signature. Therefore, in order to collect low RFI data, several trips to very remote radio quiet locations were organised.}
Much better quality data, at $\sim$270\,ms time resolution, were collected at the Eyre Bird Observatory (EBO: Fig.~\ref{fig_wa_map}).
The EBO is a very remote location on the coastline of the Southern Ocean and is hundreds of kilometres away from high power radio transmitters.
\red{Surprisingly however, we found that even in such a remote location the night time data quality was affected due to long distance RFI propagation from distant coastline transmitters in South Australia, Victoria and even Tasmania (up to 1500\,km away).
The signals were, most likely, refracted via a tropospheric ducting mechanism, which allows for long distance propagation and is common in summer and particularly severe near the coastline \citep{1457406}.}
Thus, especially after dusk there were a lot of RFI signals in the FM and digital TV bands.\footnote{We found some tropospheric ducting forecast services to be good predictors of the level of RFI seen in the data. E.g. \url{http://www.dxinfocentre.com/tropo_aus.html} }
Particularly, the later were undesired as they affected the most promising band, where the biconical antenna is best matched. 
Hence, the third site was inland at the Wondinong Station (Fig.~\ref{fig_wa_map}), where the best data to date were collected.

\subsection{RFI excision and data reduction}
The main requirement for the BIGHORNS deployment site was for a low RFI environment. Thus, the collected data are intrinsically relatively clean, even within the FM band. 
The main source of the high power RFI signals are from the low orbit telecommunication \mbox{ORBCOMM} satellites 
at around 137.5\,MHz (right-hand circularly polarised) and from airplanes in the frequency band $117.975-136$\,MHz.
The power transmitted by these devices is relatively high and can saturate the receiver.
Very high power in a narrow band can also cause distortions in the noise floor of the remaining spectrum. 
In order to efficiently exclude RFI affected data they should be collected at the highest possible time resolution, which is limited by the maximum data transmission and storage rates in the system.
Therefore, the Wondinong Station data were collected with 50\,ms resolution.

Due to the fact that the calibration switch operates autonomously, the first data processing step identifies integrations collected on the antenna and reference source and saves this information to the FITS file headers.
The state identification procedure calculates total power in every integration in a certain frequency band ($60-110$\,MHz for the Wondinong data), compares it with a threshold value
and flags as an antenna or reference integration. The threshold value is typically calculated automatically (based on the distribution of total power) or can be pre-defined according to observed total power for each of the states.
When a state transition is detected, three surrounding integrations are flagged as undefined in order to be excluded from further analysis.
In the next processing step, in order to excise data affected by RFI, the following criteria are applied:

\begin{figure}
  \begin{center}
    \leavevmode
    \includegraphics[width=3in]{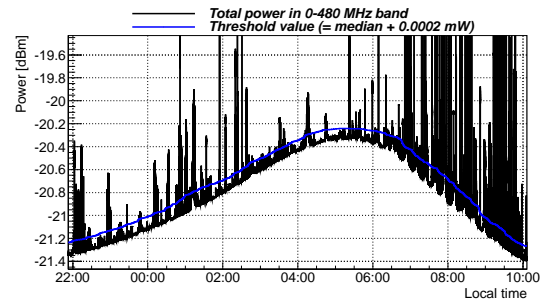}
    \caption{\red{Total power in the $0-480$\,MHz frequency band observed over 12 hours at the Wondinong Station with a cut-off threshold value.}}
    \label{fig_total_power_wond201404}
  \end{center}
\end{figure}


\begin{figure*}
  \begin{center}
    \leavevmode
	\includegraphics[width=6in]{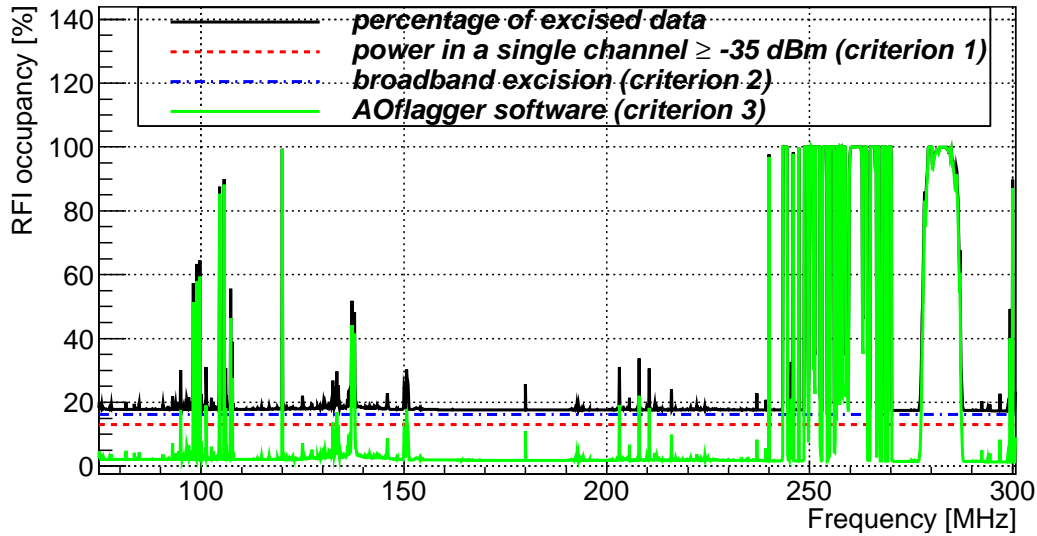}
    \caption{Percentage of all the Wondinong Station data excised by the three criteria in the frequency resolution of 117.2\,kHz. The excision rates of each individual criteria were calculated independently of the other criteria. \red{The power based criteria are constant across the frequency at $\approx$13\% (red curve) and $\approx$16\% (blue curve). The heavily RFI affected channels were almost entirely rejected by the AO-flagger software. Therefore, the green curve lies on top of the black curve in these channels.}}
    \label{fig_rfi_cuts_contrib}
  \end{center}
\end{figure*}

\begin{figure*}
  \begin{center}
    \leavevmode
	\includegraphics[width=6in]{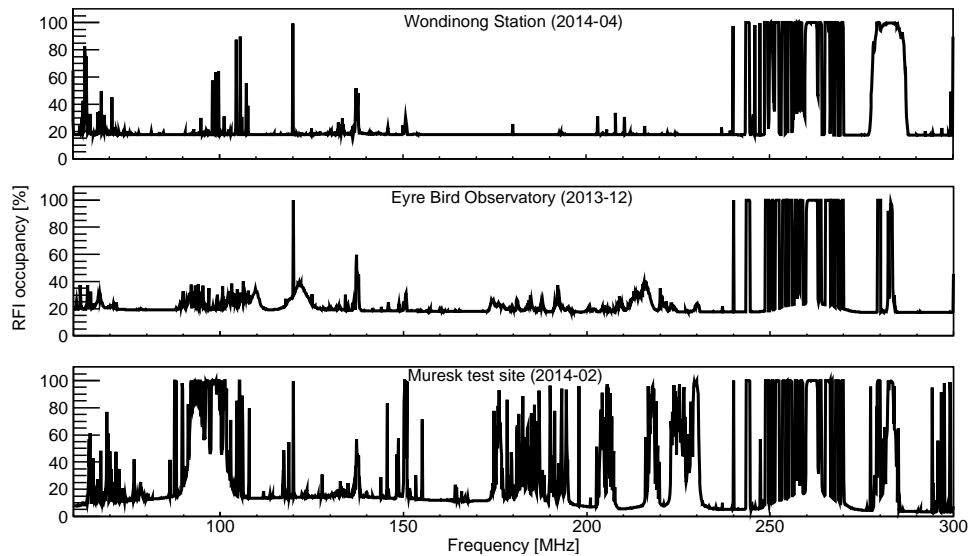}
    \caption{Comparison of RFI occupancy between the Wondinong Station, the EBO and Muresk. Please note that RFI threshold values used in criteria 1 and 2 (see text) for the EBO and Wondinong data would reject all the Muresk data. Thus, the figure shows occupancy obtained after cuts appropriate for the Muresk data were applied, which situates the ``rejection floor'' slightly below the other two, but also shows a lot of narrow band emission. 
       } 
    \label{fig_occupancy_plot_wond_ebo_muresk}
  \end{center}
\end{figure*}

\begin{figure*}
  \begin{center}
    \leavevmode
	 \includegraphics[width=6in]{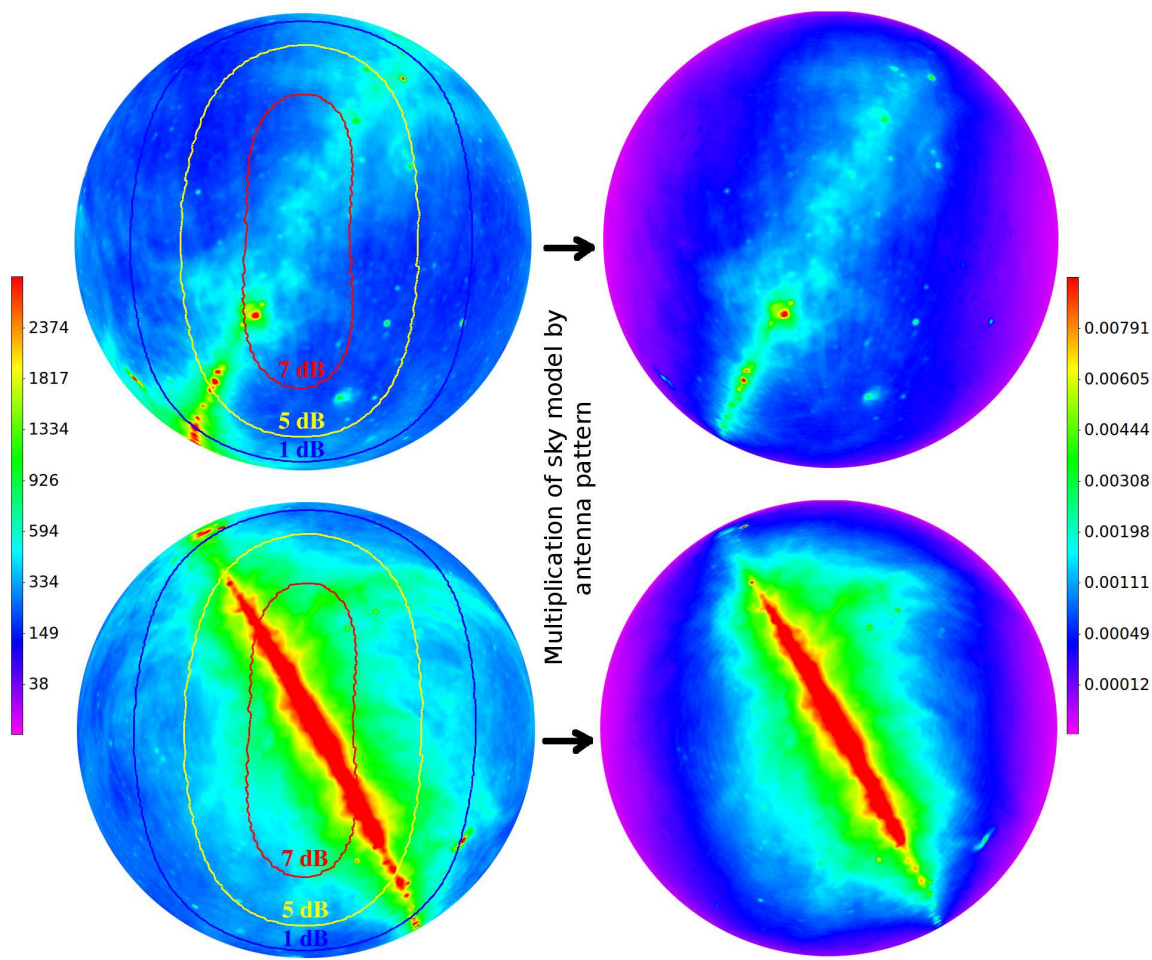}
    \caption{\red{Prediction for sky model brightness temperature $T_{\nu}(\theta,\phi)$ with over-plotted contours of simulated antenna pattern $P_{\nu}(\theta,\phi)$ (left panel) and their product (right panel) at	$\nu$=150\,MHz. 
				 The upper pair of images corresponds to Galactic Centre below the horizon (at 2014-04-06 10:34:51 UT, corresponding to approximate local sidereal time $07^h27^m$), and the lower images correspond to Galactic Centre transit (at 2014-04-06 21:07:51 UT, corresonding to approximate local sidereal time $18^h02^m$).
             Summation of values in all the pixels of the original images (before the orthographic projection) in the right panel leads to a $T_{model}$(150 MHz) as in the corresponding curve in Figure~\ref{fig_calibdata_vs_model}} }
    \label{fig_skymodel_and_antpat}
  \end{center}
\end{figure*}
\begin{enumerate}
\item \textbf{Single channel excision} - if the power measured by the spectrometer exceeded -35\,dBm in any single channel the whole integration was excised. The value of the threshold was determined empirically in the laboratory with a signal generator connected to the antenna input of the front-end. It has been observed that power $\ge$-35\,dBm affects the noise floor by $\ge$1\,\%. The threshold was set at -15\,dBm for the Muresk test data. 
                                         The average excision rate of this criterion\footnote{The average excision rates of every criteria were calculated independently of the other criteria} for the data from Wondinong Station was $\approx$13\%. 
\item \textbf{Broadband excision} - \red{if the sum of power in the entire band was above the median galactic noise level plus the threshold of -37.5\,dBm (corresponding to $5\times10^8$ in the original arbitrary power units in which the cut is performed at the early data processing stage) the integration was excised (Fig.~\ref{fig_total_power_wond201404}). For the Muresk data the threshold value was \mbox{-14.5\,dBm} ($10^{11}$ in arbitrary units).}
											The purpose of this criterion is to excise integrations affected by high RFI power appearing in more than one channel. The average excision rate of this criterion for the data from Wondinong Station was $\approx$16\% (Fig.~\ref{fig_rfi_cuts_contrib}).
\item \textbf{AO-flagger} - all the data were flagged with the AOflagger software \citep{2012A&A...539A..95O}, and the flagged data were excluded from further processing.
									 The ``rejection floor'' of this criterion is $\sim1.3-2$\% in the frequency band $60-300$\,MHz for the data from Wondinong Station and reaches 100\% for RFI affected channels (Fig.~\ref{fig_rfi_cuts_contrib}).
\end{enumerate}

The average excision rates of each criterion applied to the data collected at Wondinong Station are shown in Figure~\ref{fig_rfi_cuts_contrib}. The RFI occupancy comparing the three sites, where the system was deployed in years 2012-2014, is shown in Figure~\ref{fig_occupancy_plot_wond_ebo_muresk}.
After the RFI excision process, the data volume was reduced to a more manageable volume by averaging every N integrations on the antenna and a reference load, where N was 700 for the sky signal and varied between 200 and 300 on the calibration signal.
The number of averaged single integrations in every frequency channel was saved to a separate FITS file for later analysis. Further data analysis was performed on the reduced dataset.

\subsection{Sky signal modelling}
In order to calculate the expected sky spectrum, a sky model \citep{2008MNRAS.388..247D} was integrated with a simulated antenna pattern.
A model of the biconical antenna 52\,cm above a $3\times3$\,m ground screen and soil was developed in the FEKO 6.3 package in order to calculate the expected antenna pattern, $P_{\nu}(\theta,\phi)$, and efficiency, $\eta(\nu)$.
The coordinates $(\theta,\phi)$ are angles in the antenna frame and can be directly translated to horizontal coordinates (az,el) according to orientation of the antenna in the field.
For a given time and frequency, $\nu$, the antenna pattern $P_{\nu}(\theta,\phi)$ was integrated over the entire sky with the sky brightness temperature $T_{\nu}(\theta,\phi)$, obtained from the sky model, according to the formula:

\begin{equation}
T_{pattern}(\nu) = \frac{\int_{4\pi} P_{\nu}(\theta,\phi) T_{\nu}(\theta,\phi) d\Omega}{\int_{4\pi} P_{\nu}(\theta,\phi) d\Omega}.
\label{eq_sky_integration}
\end{equation}

The above formula is time dependent through the implicit time dependence of $T_{\nu}(\theta,\phi)$. 
Many equations and quantities in this paper are frequency and time dependent, but their frequency and time dependence is usually not explicitly shown for brevity.

\red{Predictions for the sky model, antenna pattern, and their product in zenith orthographic projection are shown in Figure~\ref{fig_skymodel_and_antpat}.} 

\begin{figure}
  \begin{center}
    \leavevmode
	 \includegraphics[width=3in]{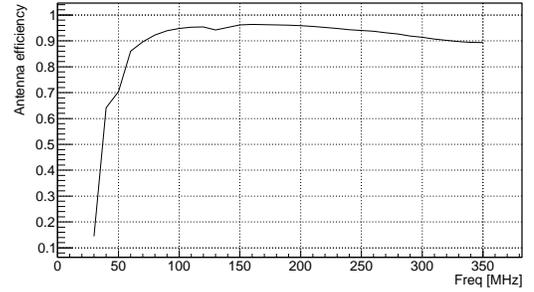}
    \caption{Efficiency ($\eta$) of the biconical antenna obtained from the FEKO model. The dip at $\approx$130\,MHz is caused by ground screen related effects, and it is also observed in the data, although slightly shifted in frequency (Fig.~\ref{fig_calibdata_vs_model}).
    Using the FEKO model it was verified that the depth of dip depends on the height of the antenna above the ground screen and that the dip is not present when the antenna is in free space (without the ground screen).}
    \label{fig_bicon_eff_wondinong}
  \end{center}
\end{figure}

Resistive losses in the antenna rods and due to signal picked up from the ground were estimated in the FEKO simulation (Fig.~\ref{fig_bicon_eff_wondinong}). Thus, the expected output signal from the antenna (before the balun) is
\begin{equation}
T_{model}(\nu) = T_{pattern}(\nu) \eta(\nu) + (1 - \eta(\nu)) T_{amb}, 
\end{equation}
\red{where $T_{amb}$ is ambient temperature}. Further losses in the signal chain (before the RF-switch inside the front-end), due to balun and cable inside the antenna, were included in the calibration process.
A comparison of calibrated and modelled spectra of the sky at Wondinong Station when the Galactic Centre was transiting and when it was below the horizon are shown in Figure~\ref{fig_calibdata_vs_model}. 

%

\subsection{Absolute calibration of the signal}
\label{sec_abolute_calib}

In the existing calibration schema a two position switch was used, switching between the antenna and a 50\,$\Omega$ resistor at ambient temperature (cold load).
The power spectra observed when the switch is set to antenna and load can be expressed by the following equations:

\begin{equation}
P_{ref} = g  (T_{amb} (1 - |\Gamma_{l}|^2) + T^{r}_{rcv} )
\label{eq_power_ref}
\end{equation}

\begin{equation}
P_{ant} = g  T_{rec} 
\label{eq_power_ant}
\end{equation}

\red{where $P_{ref}$ and $P_{ant}$ are the powers measured on the antenna and reference respectively, and g is the system gain post low noise amplifier (LNA) at the receiver. }

\red{$T^{r}_{rcv}$ is the receiver noise temperature when the RF-switch is set to the 50\,$\Omega$ terminator measured with the VNA between points \textbf{L} and \textbf{D}
(Fig.~\ref{fig_ebo_setup}), and $|\Gamma_{l}|^2$ is the reflection coefficient between the receiver and reference source (in this case a 50\,$\Omega$ resistor).}

\red{$T_{rec}$ is the sum of signal collected by the antenna ($T_{ant}$) and transmitted to the receiver (not reflected back) and full (including noise reflections in the antenna-receiver system) receiver noise ($T_{rcvn}$): }

\begin{equation}
T_{rec} = T_{ant} (1 - |\Gamma_{al}|^2)  L + T_{rcvn},
\label{eq_trec}
\end{equation}

\red{where L is a frequency dependent loss between the antenna terminals and the first LNA (due to losses in the RF-switch and an optional attenuator). Losses of these components as a function of frequency were measured with the VNA.}
\red{$|\Gamma_{al}|^2$ is the reflection coefficient between the antenna and receiver. Hence, the term $(1 - |\Gamma_{al}|^2)$ is the fraction of antenna signal entering the receiver (not reflected back to the antenna).}
\red{$\Gamma_{al}$ can be expressed in terms of impedances of the antenna ($Z_a$) and receiver ($Z_l$) as:}

\begin{equation}
\Gamma_{al} = \frac{Z_a - Z^{*}_l}{Z_a + Z_l}.
\label{eq_gamma_al}
\end{equation}

\red{$T_{ant}$ in equation~\ref{eq_trec} is the antenna temperature (in lossless antenna case $T_{ant} \approx T_{sky}$). 
Because frequency dependance of losses inside the off-the-shelf biconical antenna could only be estimated according to datasheets of the balun and cable, non-zero losses inside the antenna ($L_a \ne 0)$ were optional in the calibration process as they might be a source of significant uncertainty.
After taking into account non-zero losses inside the antenna, $T_{ant}$ can be expressed as:}

\begin{equation}
T_{ant} = T_{sky}L_a + (1-L_a)T_{amb}.
\label{eq_tant}
\end{equation}

\red{Finally, the receiver noise ($T_{rcvn}$) in equation~\ref{eq_trec} can be expressed as a sum of the standard \footnote{\red{We refer to standard receiver noise temperature as measured with the VNA which uses a well matched reference source. In contrast to real, mismatched sources such as an antenna.}} receiver noise temperature $T^{a}_{rcv}$ (measured between points \textbf{I} and \textbf{D} in Figure~\ref{fig_ebo_setup}), and a reflected noise term ($T_{nw}$) due to receiver input noise reflected back and forth between the antenna and receiver (``noise waves'') resulting from the impedance mismatch of these elements (Fig.~\ref{fig_ebo_setup}):}

\begin{equation}
T_{rcvn} = T^{a}_{rcv} + T_{nw}.
\label{eq_trcvn}
\end{equation}

\red{In the first approximation $T_{rcvn} \approx T^{a}_{rcv}$, but higher accuracy calibration requires taking into account reflected noise terms ($T_{nw}$). The biconical antenna 52\,cm above the ground screen (Section~\ref{sec_instrument_desc}) is very well matched to the front-end \mbox{($|\Gamma_{al}|^2<$~0.01)} only in the frequency band $\sim$172-184\,MHz. Thus, the contribution of reflected noise is substantial outside this range.}
The contribution of the reflected noise was estimated with an approach described by \citet*{2012RaSc...47.0K06R}, where it is treated as noise waves due to LNA input noise reflected back and forth between the antenna and LNA. 
\red{The noise wave amplitudes were suppressed by a 3~dB attenuator inserted between the antenna and input of the front-end. However, this is undesirable because it significantly increases the noise temperature of the receiver (improvements in the presented system will be discussed in Section~\ref{sec_future_improvements}).}
The procedure of determining noise wave contribution was described in detail in the aforementioned paper.
In summary, the amplitudes of the noise wave terms were estimated by measuring a power spectrum of an open (or shorted) cable of sufficient length (in our case $\sim 10$\,m) connected to the input of the front-end (instead of the antenna); and measuring the reflection coefficient of this cable; and the reflection coefficient of the front-end (which was measured anyway).
The observed power spectrum of the cable can be described by equation~\ref{eq_trec}, where $T_{ant}$ is set to $T_{amb}$ (as the cable ``sees'' the physical temperature), and $T_{rcvn}$ can be expressed in terms of correlated ($T_c,T_s$) and uncorrelated ($T_u$) portions of the noise from the input of the LNA as in equation 8 in \citet*{2012RaSc...47.0K06R}.
After assuming certain parameterisation (e.g. linear) of the $T_c,T_s$ and $T_u$ terms, their parameters can be fitted in order to reproduce the observed power spectrum of the cable.

\red{An alternative, anechoic chamber based method of determining full receiver noise (including reflected noise contribution) was also developed and tested.}
It is based on the assumption that the brightness temperature in the \red{shielded ($\sim$80~dB at the frequencies of interest)} anechoic chamber is equal to the ambient temperature.
\red{A proper analysis of how well the anechoic chamber approximates a black-body is considered as very important if the method is going to be further used and improved, but is beyond the scope of this paper. However, a first order approximation can be obtained based on normally incident plane wave and reflectivity (deviation from a perfect absorber) of the absorber (FS-600H), which is $\sim$10\% at 20\,MHz, $\sim$3\% at 30\,MHz and $\sim$1\% above 80\,MHz. Thus, we estimate that the chamber acts as a black-body within a few percent down to $\sim$30\,MHz and even better (within $\sim$1\%) at frequencies above 80\,MHz.}
Equation \ref{eq_trec} can then be converted to:

\begin{equation}
T_{rcvn} = T_{rec} - T_{amb} (1 - |\Gamma_{al}|^2)  L , 
\label{eq_noise_waves_chamber}
\end{equation}

where all the values on the right hand side can be determined: $T_{rec}$ according to equation~\ref{eq_trec2}; $T_{amb}$ with a temperature probe; and coefficient $\Gamma_{al}$ can be calculated from equation \ref{eq_gamma_al} after measuring 
the reflection coefficient of the antenna in the anechoic chamber. In order to measure $T_{rcvn}$ the system was deployed in the anechoic chamber in almost exactly the same form as in the field.
The resulting values of $T_{rcvn}$ were later used to calibrate the sky data.
The method yields very similar results to the method of \citet*{2012RaSc...47.0K06R} and might be further improved and used in the future.

Equations \ref{eq_power_ant}-\ref{eq_trcvn} can be solved to determine $T_{rec}$ which can be expressed as

\begin{equation}
T_{rec} =  \frac{P_{ant}}{P_{ref}} ( T_{amb} (1 - |\Gamma_{l}|^2) + T^{r}_{rcv} ).
\label{eq_trec2}
\end{equation}

The factor $(1 - |\Gamma_{l}|^2)$ can be approximated by 1, because a 50\,$\Omega$ terminator is very well matched to the front-end input.
Finally, the calibrated sky temperature, corrected for reflections in the antenna-LNA system, can be obtained from equations \ref{eq_trec}, \ref{eq_tant}, \ref{eq_trcvn} and \ref{eq_trec2} as :

\begin{equation}
T_{sky} = \frac{T_{rec} - T_{rcvn}}{(1 - |\Gamma_{al}|^2) L L_a} + (1 - \frac{1}{L_a})T_{amb} .
\end{equation}

\subsection{Statistical errors}
\label{subsec_stat_errors}

The statistical error on $T_{rec}$ can be expressed by the following formula:

\begin{equation}
\begin{split}
\delta T_{rec} = \frac{P_{ant}}{P_{ref}} \times \; \; \; \; \; \; \; \; \; \; \; \; \; \; \; \; \; \; \; \; \; \; \; \; \; \; \; \; \; \;\\ 
\sqrt{ ( T_{amb} + T^{r}_{rcv} )^2 \bigg( \Big(\frac{\delta P_{ref}}{P_{ref}} \Big)^2 + \Big(\frac{\delta P_{ant}}{P_{ant}}\Big)^2 \bigg) + \delta T_{amb}^2 } \;,
\end{split}
\label{eq_trec_err}
\end{equation}

\red{where $\delta T_{amb}$ is the error on the temperature measurement and $\delta P_{ref}$ and $\delta P_{ant}$ are statistical errors of the power measured on the reference and antenna, respectively.}
\rednew{Equation \ref{eq_trec_err} holds under the assumption that the errors ($\delta P_{ref}$, $\delta P_{ant}$ and $\delta T_{amb}$) are independent and Gaussian distributed, which we verified on subsets of the data.}
It can be simplified under the assumption that the error on the measured power satisfies the radiometer equation $\delta P/P = \sqrt{B \tau}$, where B is the resolution of the frequency bin and $\tau$ is the integration time.
If a long integration time $\tau$ is achieved by averaging M single integrations of duration $\tau_i$ then the error on the temperature measurement also decreases as $\delta T_{amb} = \sigma_{amb}/\sqrt{M} = \sigma_{amb}/\sqrt{\tau/\tau_i}$, 
where $\sigma_{amb}\approx$0.03\,K is the estimated error on a single measurement with the presently used temperature probe.
Consequently, the error on such an averaged $T_{rec}$ can be expressed by
\begin{equation}
\delta T_{rec} = \frac{P_{ant}}{P_{ref}} \sqrt{ \frac{( T_{amb} + T^{r}_{rcv} )^2}{B}\Big(\frac{1}{\tau_{ant}} + \frac{1}{\tau_{ref}}\Big)  + \sigma_{amb}^2\frac{\tau_i}{\tau} } \;,
\label{eq_trec_err_simplified}
\end{equation}

where $\tau_{ant}$ and $\tau_{ref}$ are integration times on antenna and reference, respectively (in a situation when they are not the same).

\subsection{Comparison of calibrated data and sky model predictions}
\label{subsec_calib_vs_skymodel}

\begin{figure}
  \begin{center}
    \leavevmode
	  \includegraphics[width=3in]{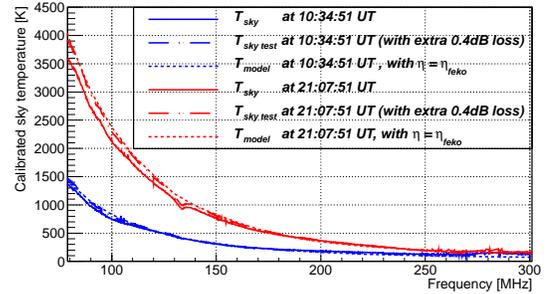}
     \caption{Calibrated data collected on 2014-04-06 at the Wondinong Station (with RFI affected channels removed) compared with the sky model integrated with antenna pattern. The dash-dotted lines show a calibration with an extra 0.4\,dB loss inside the antenna, which could be the case due to lack of exact characteristics of the antenna.}
    \label{fig_calibdata_vs_model}
  \end{center}  
\end{figure}

An example of two calibrated integrations (average of 700 single 50\,ms integrations) when the Galaxy was transiting and below the horizon is shown in Figure~\ref{fig_calibdata_vs_model}.
The relative difference between calibrated spectrum and the sky model is below 10\% in the $70-200$\,MHz band, and gets down to 5\% when \red{an extra 0.4\,dB of unaccounted attenuation is added to the attenuation inside the antenna}.
These values are within the sky model error, which was estimated by its authors \citep{2008MNRAS.388..247D} to be $\approx$10\% at low frequencies ($20-400$\,MHz).

\begin{figure}
  \begin{center}
    \leavevmode
	  \includegraphics[width=3in]{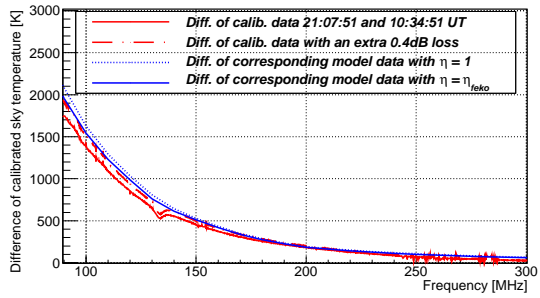}	  	
     \caption{The difference of the calibrated spectrum at two different epochs (21:07:51 and 10:34:51 2014-04-06 UT) compared with the corresponding differences of the sky model. The red dash-dotted line corresponds to difference of calibrated data with an extra 0.4\,dB (Fig.~\ref{fig_calibdata_vs_model}). The two model curves were generated for antenna efficiency $\eta$=1 and $\eta$ as from the FEKO simulation. }
    \label{fig_calib_difference_vs_model}
  \end{center}  
\end{figure}

The discrepancy may be attributed to an overestimation of the antenna efficiency ($\eta$), underestimation of the signal loss between the antenna and the LNA, incorrect subtraction of reflected noise contribution, or uncertainty in the sky model itself.
In order to test our understanding of the signal calibration without relying on the exact knowledge of the additive noise terms $T_{rcvn}$ (\red{mainly uncertainty of reflected noise}), the difference of the two calibrated integrations was compared with a difference of the corresponding sky model predictions (Fig.~\ref{fig_calib_difference_vs_model}).
Figures \ref{fig_calibdata_vs_model} and \ref{fig_calib_difference_vs_model} show that an accurate absolute calibration requires a very good knowledge of the antenna efficiency, which is a poorly constrained quantity. 
In the presented analysis the value of $\eta$ was obtained from the FEKO simulation, and in practise it is difficult to have a better estimation.
\red{The antenna efficiency is typically measured with a $\sim$20\% precision, 5-10\% is considered good and the best accuracy that can be achieved is $\sim$1\% \citep{wheeler_cap}.
Even such an extremally good measurement would introduce a frequency dependent error $\gtrsim$1\,K outside $\sim130-170$\,MHz band (at $\sim$150\,MHz ``cold sky'' temperature equals typical ambient temperature thus antenna efficiency does not matter).
Although it might not be possible to absolutely calibrate the signal to a mK precision, as long as the antenna efficiency is a smooth and ``well behaved'' function (without bumps and wiggles) of frequency, the global EoR signature can still be detected.
The precise absolute calibration also necessitates good characteristics of losses inside the antenna which, in the case of the off-the-shelf biconical antenna, could not be measured precisely (without breaking the antenna) and only approximate data-sheet values were used.}

\subsection{Data quality assessment}
\begin{figure}
  \begin{center}
    \leavevmode
	 \includegraphics[width=3in]{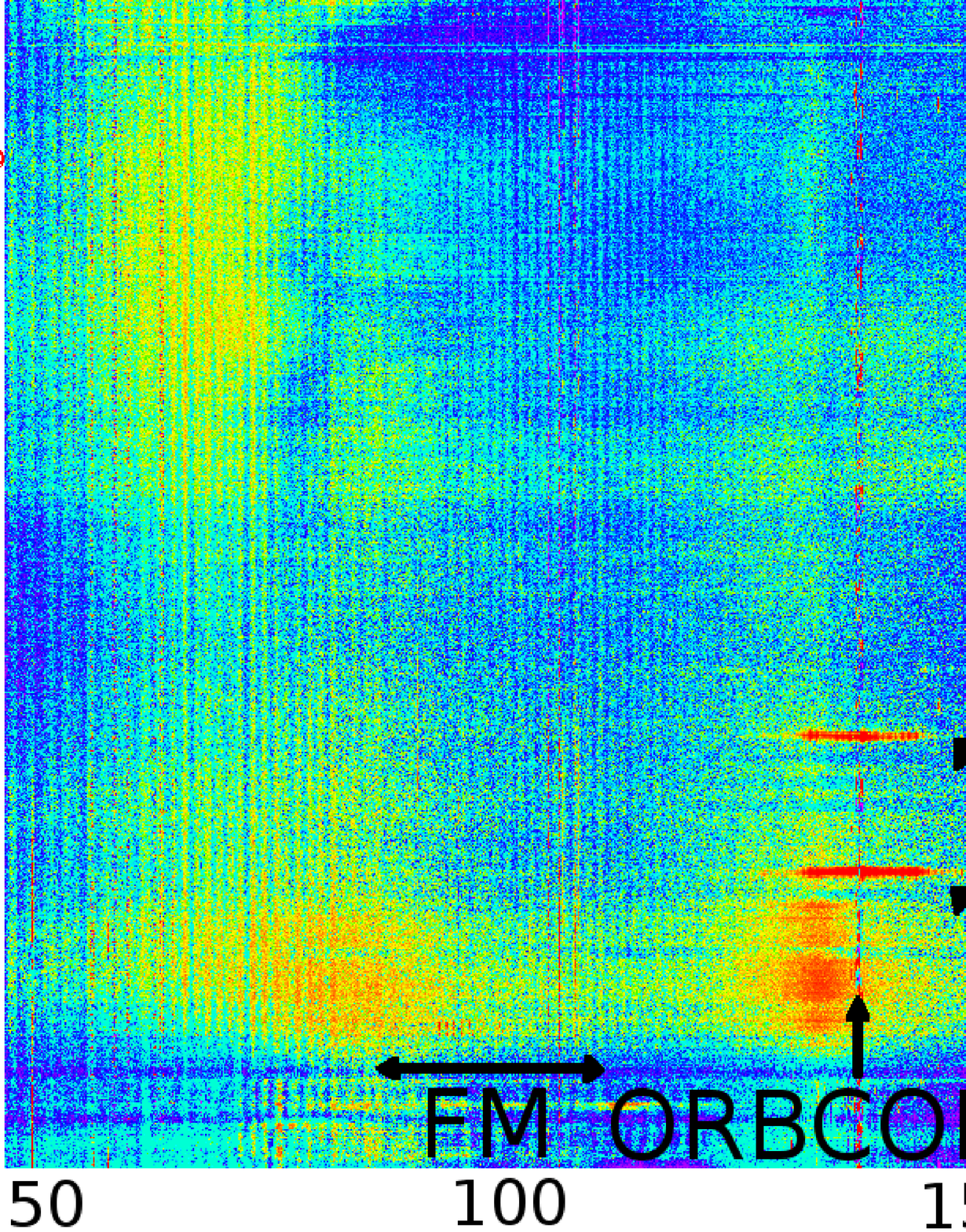}
     \caption{\red{Normalised dynamic spectrum of approximately 66~hours. Frequency scale is on horizontal axis and time flows from the bottom to the top. The first night was affected by thunderstorms within some 100~km. Nevertheless, excluding frequency bands affected by RFI (marked in the figure) due to FM (87.5-108~MHz) and ORBCOMM (137-138~MHz and sometimes nearby channels), the night time data is mostly stable to $\lesssim$1\% in the frequency band 50-235\,MHz. Several other features such as: daytime solar activity and RFI due to military satellites ($\sim$235-290~MHz) were also marked in the image. The $\approx$1.275\,MHz ripple is well visible as vertical lines at frequencies below 100\,MHz.}}
    \label{fig_ratio_of_dynamic_spectrum}
  \end{center}  
\end{figure}

The calibration schema was used to calibrate 5 days of averaged antenna integrations by the corresponding averaged reference integrations and saving the resulting dynamic spectra to FITS files.
In order to assess the quality of the antenna data, a median dynamic spectrum for 1 sidereal day was generated by grouping all data into local sidereal time (LST) bins of size approximately 1 min and taking the median of each bin.
Every integration of the entire 5 day dataset was then divided by the corresponding value (with the closest LST) of the median dynamic spectrum.
An example piece of the normalised dynamic spectrum is shown in Figure~\ref{fig_ratio_of_dynamic_spectrum}.
\red{The daytime data are significantly affected by solar activity, but the night time data are stable (typically to less than $\sim$1\%). 
Several RFI related features marked in the image are due to: military satellites at $\sim$235-290~MHz, FM band at 87.5-108~MHz and ORBCOMM satellites at 137-138~MHz. 
The relatively high power transmitted by ORBCOMM satellites sometimes caused some undesired leakage-like effects on the nearby channels (marked with arrows in the bottom of the plot), which were not excised by the earlier RFI flagging criteria, but were identified and excluded based on the normalised dynamic spectrum. 
The first 24~hours were also affected by thunderstorms within some 100~km. 
The $\approx$1.275\,MHz ripple resulting from the reflections in the 100~m cable is present almost in the entire band, but it is well visible as vertical lines below 100\,MHz due to a particularly bad match at these frequencies.
Some horizontal lines are due to ORBCOMM power affecting the entire band which may enforce an even more rigorous threshold in the RFI excision criteria 2.
The normalised dynamic spectrum is a very useful diagnostic tool which enables identification and excision from further analysis integrations still affected by RFI or other undesired effects.
}

\section{Limitations of the system}
\label{sec_system_limitations}

Averaging of the spectrum to mK precision requires collecting data over several days, as only a few hours of the best data per day can be used. Such a procedure requires a stable system. 
Stability of the BIGHORNS system was tested in the laboratory and in the anechoic chamber.

\subsection{Stability of the reference data}
During the laboratory tests the antenna input of the front-end was terminated with a 50\,$\Omega$ resistor. 
A large laboratory dataset (475\,hours) was reduced by averaging all 200 integrations ($\sim$270\,ms each) in every single FITS file ($\sim$54.6\,sec) and saving them to a separate output FITS file.
The variations of the recorded signal were estimated by normalising the averaged dynamic spectrum to the first or a median integration.
The response of the system changes by a few percent due to temperature variations and changes in the system's gain as a function of temperature.
Switching between the antenna and a reference source allows us to calculate the instantaneous gain and thus eliminate the effects of gain variations from the calibrated data. 

\begin{figure}  
  \begin{center}
   \leavevmode
   \includegraphics[width=3in]{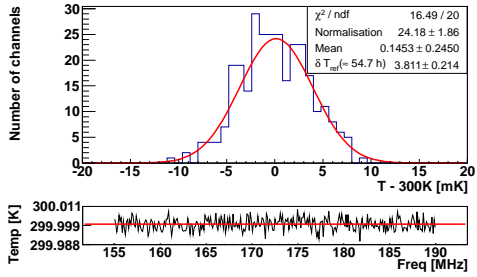}
   \caption{\red{Fit of constant temperature in the 155-190~MHz band to calibrated reference integration of $\approx$54.7~hours duration (lower panel), and a distribution of residuals with a Gaussian fit superimposed (upper panel).
   The standard deviation of the fitted Gaussian represents $\delta T_{ref}(\tau \approx$54.7~hours$) \approx 3.8$~mK in Figure~\ref{fig_xmass_ref_rms_vs_time}.}}
	\label{fig_tconstfit_and_distrib}
   \end{center}
\end{figure}

\begin{figure}
  \begin{center}
    \leavevmode
   \includegraphics[width=3in]{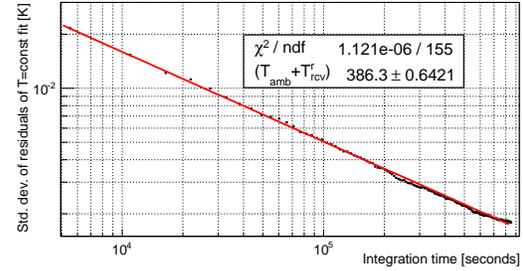} 
    \caption{The error $\delta T_{ref}(\tau)$ in a single frequency bin of \red{117.2\,kHz within 155-190~MHz frequency band (derived as described in the text) as a function of integration time $\tau$. Every point in this figure was obtained from the Gaussian fit to a distribution of residuals (Fig.~\ref{fig_tconstfit_and_distrib}).}}
    \label{fig_xmass_ref_rms_vs_time}
  \end{center}
\end{figure}

Using the same large laboratory dataset, the ability of the system to integrate down according to the radiometer equation was tested. 
Because spectra collected on the terminated antenna input are indistinguishable from those collected on the reference, odd integrations were calibrated with the even ones.
Such calibrated data were later averaged and a constant room temperature was fitted in the 155-190\,MHz sub-band to the resulting average spectrum.
The Gaussian distribution of residuals of this fit yielded a statistical error $\delta T_{ref}\approx$1.7\,mK of the mean integration resulting from averaging all the integrations, which is within 0.1\,mK of what is expected from equation \ref{eq_trec_err_simplified} and $\tau\approx$ (475/2) hours and B=117.2\,kHz.
\red{The same procedure of fitting Gaussian to a distribution of residuals (Fig.~\ref{fig_tconstfit_and_distrib}) led to determination of the error as a function of integration time which is shown in Figure~\ref{fig_xmass_ref_rms_vs_time}.}

The stability of the data collected at Wondinong Station was also assessed in the same way.
Every 200 reference integrations were averaged to a single ($\approx$10\,sec) integration, and odd integrations were calibrated by the even ones.
All the calibrated reference integrations ($\approx$13.2\,hours) were averaged and a constant (across frequency) temperature fit was subtracted. 
The error of $\approx7.4$\,mK was derived from the distribution of residuals and agrees very well with a value $\approx7.3$\,mK obtained from equation \ref{eq_trec_err_simplified} for a given integration time and $(T_{amb}+T^{r}_{rcv})$=386\,K.
The dependence on the integration time is also in a good agreement with the expectations.

Finally, the quality of the reference data was always verified by normalising the averaged dynamic spectrum of the reduced reference data to the first (or median) integration, which would allow identification of any undesired variations in the reference signal and system's response.



\subsection{Stability of the antenna data}

\begin{figure}
  \begin{center}
    \leavevmode
	  \includegraphics[width=3in]{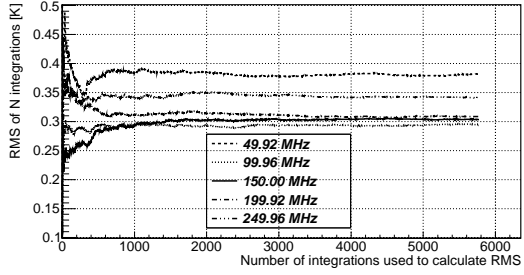}
     \caption{Standard deviation of the $T_{rec}$ as a function of number of sample integrations used to calculate it for selected frequency channels.}
    \label{fig_rms_vs_nint}
  \end{center}  
\end{figure}

In order to test the stability of the antenna data, the system was deployed in the anechoic chamber in almost the same form as in the field, and a few days worth of data were collected.
The uncalibrated antenna and reference power varies by $\sim$1\%. The calibrated signal varied by $\sim$0.1\% which corresponds to variations of the physical temperature in the chamber.
The data were collected at $\sim$270\,ms resolution and were reduced as described in Section~\ref{sec_data_processing}.
Single integrations were averaged resulting in a reduced dynamic spectrum at $\sim$54.6\,sec resolution (200 single integrations) and reference at $\sim$15.8\,sec (43-65 single integrations) resolution.
The resulting dynamic spectrum was calibrated to $T_{rec}$ according to equation ~\ref{eq_trec}.
Then, for every frequency channel a standard deviation of $T_{rec}$ was calculated from a sample of increasing number of integrations (Fig.~\ref{fig_rms_vs_nint}).

\begin{figure}
  \begin{center}
    \leavevmode
	 \includegraphics[width=3in]{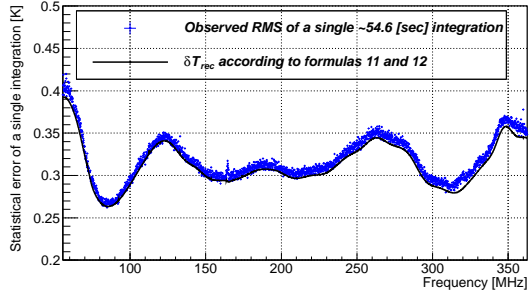}
    \caption{Standard deviation of $T_{rec}$ calculated for all integrations of the chamber data (Fig.~\ref{fig_rms_vs_nint}) compared with the expected value of $\delta T_{rec}$ calculated according to equation \ref{eq_trec_err_simplified}. }
    \label{fig_delta_trec_exp_vs_obs}
  \end{center}
\end{figure}

The standard deviation converges to a constant value after a sufficient number of integrations and remains constant at least over a period of $\sim$5 days of the test duration.
Its value, representing observed $\delta T_{rec}$ of a single frequency channel of a single calibrated mean integration ($\tau_i\approx$54.6\,sec), varies across frequency between 0.25-0.4~K.
The expected values of $\delta T_{rec}$ were calculated according to equation~\ref{eq_trec_err_simplified}, and a good agreement between the observed and expected values of $\delta T_{rec}$ is shown in Fig.~\ref{fig_delta_trec_exp_vs_obs}.
The good understanding of the statistical error on the calibrated data collected in the anechoic chamber enabled us to estimate expected statistical errors for the sky data.

\begin{figure}
  \begin{center}
    \leavevmode
	 \includegraphics[width=3in]{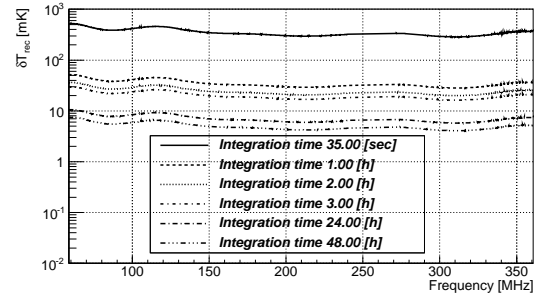}
    \caption{The expected statistical error $\delta T_{rec}$ calculated according to equation \ref{eq_trec_err_simplified} as described in the text for the case of ``cold sky'' (when the Galactic Centre was in the lowest point below the horizon).
				\red{The error does not significantly increase at lower frequencies, as could be expected from the high sky temperatures (Fig.~\ref{fig_expected_global_eor}), due to the very poor antenna match at these frequencies (reflecting back more than half of the sky signal).}}
    \label{fig_delta_trec_cold_sky}
  \end{center}
\end{figure}

The expected $\delta T_{rec}$ for the sky data was calculated according to equation~\ref{eq_trec_err_simplified} where the ratio $P_{ant}/P_{ref}$ for a ``cold sky'' was taken from the Wondinong data.
The values of $\delta T_{rec}$ for single 117.2\,kHz channels calculated for several integration times (including a single average integration of $\sim$35\,sec) are shown in Fig.~\ref{fig_delta_trec_cold_sky}. 
The expected error of 10 average channels, corresponding to a frequency resolution of $\sim$1.17\,MHz, would be $\sqrt{10}$ smaller.

\subsection{Limitations of calibration}
\label{subsec_calib_limitations}



The two positional calibration schema works under the assumption that noise and S-parameters characteristics of the LNA do not vary significantly with variations of environmental parameters. Modern LNAs satisfy this assumption fairly well. 
The LNA response was tested in the laboratory by heating up (to $\sim$320-330\,K) and cooling down (to $\sim$276-283\,K) the first amplifier in the chain and observing changes in measured noise temperature and S-parameters.
The maximum differences of the LNA's parameters between cold and hot states were: gain $\approx$0.05\,dB; reflection coefficient $\approx$0.5\,dB; and noise temperature $\approx$1-2\,K.
The parameter changes were affected by a systematic shift (without any significant frequency structure), which could result in a systematic error of the absolute calibration $\approx$1-2\,K.
Daily variations of temperature at Wondinong Station were $\sim$288-316\,K over the period of data collection, and only $\sim$294-306\,K over 4 hours after the sunset (before the rise of the Galactic Centre).
Thus, the error on the absolute calibration was likely smaller than the estimations from the laboratory tests. Particularly, for the best parts of the data (a few hours after sunset) when the temperature was close to room temperature at which LNA's characteristics was measured in the laboratory.
\red{A new front-end equipped with an in-built ``hot and cold'' reference source will enable us to continuously determine the instantaneous receiver noise (Section~\ref{subsec_receiver_improvements}).}

\begin{figure}
  \begin{center}
    \leavevmode
     \includegraphics[width=3in]{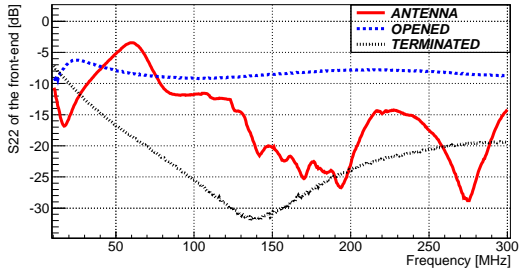}
     \caption{\red{The reflection coefficient (S22) of the front-end measured at point \textbf{O} (Fig.~\ref{fig_ebo_setup}) with the three different sources connected to its input.}}
    \label{fig_s22_fe}
  \end{center}
\end{figure}

\red{
The main difficulty in calibrating the data collected with the presented system comes from the effects of source impedance (antenna or 50\,$\Omega$ terminator) propagating further downstream the signal path.
Particularly, the aforementioned $\approx$1.275\,MHz quasi-periodic ripple due to reflections of the signal on both ends of the 100m cable connecting the front-end with second gain stage cannot be fully calibrated out and can only be modelled.
This is due to the fact that it has a different amplitude and phase as a function of frequency when the RF-switch is set to the antenna and reference. 
These effects can be diagnosed and quantified by measuring the reflection coefficient at the output of the front-end (S22) at the reference point \textbf{O} (Fig.~\ref{fig_ebo_setup}) which is significantly different when the RF-switch is set to antenna and a 50$\Omega$ terminator (Fig.~\ref{fig_s22_fe}).
Several modifications in the front-end, which should efficiently eliminate this problem, have already been tested and will be discussed in Section~\ref{subsec_receiver_improvements}.
}

\subsection{Antenna mismatch}
\label{subsec_antenna_mismatch}


\red{
The off-the-shelf biconical antenna, as presented in Section~\ref{subsec_comp_char}, is very well matched (reflection coefficient \mbox{$\lesssim$~-20\,dB}) over a relatively narrow frequency band ($\sim$172-184\,MHz).
Nonetheless, very good match in a wide frequency band ($\sim$50-100\,MHz) facilitates the calibration procedure and thus is a very important factor for the global EoR experiment.
A better matched antenna together with the ``isolation'' in the front-end (Section~\ref{subsec_calib_limitations}) can further reduce the downstream propagation effects. 
While the effects of the LNA input noise reflected between the antenna and LNA cannot be entirely eliminated, they can be significantly reduced with a very well matched antenna. 
Assuming that all the mismatch effects downstream from the first LNA can be calibrated out by cold or hot-cold reference (discussed in Section~\ref{subsec_calib_limitations}),
some requirements for optimal reflection coefficients come from the considerations of noise reflected in the antenna-receiver system and absolute calibration (Section ~\ref{sec_abolute_calib}).
In the first approximation (single reflection of the receiver input noise from the antenna) the amplitude of these reflections is proportional to the product of the antenna reflection coefficient ($|\Gamma_a|^2$) and input noise temperature of the receiver (typical value of the order of $T_{in}\sim$50\,K).
The requirement for \mbox{$T_{in}|\Gamma_a|^2$$\le$1\,mK} leads to \mbox{$|\Gamma_a|^2 \le$-47\,dBm} - nearly a perfect match. Such a good match cannot be achieved in practise over a finite bandwidth.
In the antenna engineering a \mbox{$\le$-10\,dB} match is considered good, \mbox{$\le$-15\,dB} match very good and \mbox{$\le$-20\,dB} in a wide frequency band (e.g. 100\,MHz) is extremally good and difficult to achieve.
Usually a $\le$-30dB match can be achieved in a narrow frequency band of the order of 5-10\,MHz (e.g. biconical antenna in Figure~\ref{fig_antenna_s11}). 
The condition might be relaxed by requiring only smoothness of the reflection coefficient in a restricted frequency range (tens of MHz) to satisfy the condition \mbox{$T_{in}|\Delta \Gamma_a|^2$$\le$1\,mK}, which results in condition \mbox{$|\Delta \Gamma_a|^2 \le 2 \times 10^{-5}$}. 
This means that the smoothness of the reflection coefficient itself should be smaller than $10^{-5}$, which is also difficult to achieve in practise.
Thus, either LNA input noise has to be significantly suppressed or noise reflections between the antenna and receiver have to be understood and parametrised with a required precision.
Nevertheless, a better match implies easier parametrisation of the reflections in the antenna-receiver system. Hence, it is important to have these components matched as well as possible and thus an upgrade to a better matched antenna (Section~\ref{subsec_antenna}) is one of the major modifications needed in the presented system.
}

\section{Future improvements in the system}
\label{sec_future_improvements}

\red{Based on experiences with the prototype system we identified several weak points and used this knowledge to improve the system.
The modifications described in this section have been tested in the laboratory, and a field deployment of the system implementing all the described changes is planned in the future.}


\subsection{Modifications in the front-end}
\label{subsec_receiver_improvements}

\begin{figure}
  \begin{center}
    \leavevmode
     \includegraphics[width=3in]{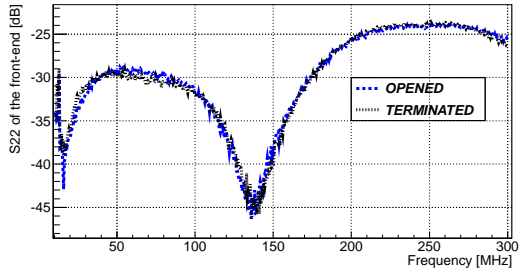}
     \caption{\red{The reflection coefficient (S22) of the new front-end measured at the output with terminated or opened input port, which are the two ``extremes'' and the antenna is somewhere in between. In this case the match is very good and almost exactly the same in both cases.}}
    \label{fig_s22_morgan_fe}
  \end{center}
\end{figure}

\red{
It has been identified that reflection coefficient of the output (S22) of a single LNA (ZX60-33LN-S+) is significantly affected by the source impedance connected to its input (Sec.~\ref{subsec_calib_limitations}).
Therefore, in order to suppress effects of the source impedance changing the reflection coefficient of the output of the front-end (downstream propagation effects) more attenuators and amplifiers were added into a modified front-end, which introduce a desired ``impedance isolation'' between its input and output.
Instead of having a single LNA, a front-end with two LNAs (ZX60-33LN-S+) providing $\sim$44\,dB of gain separated by a 10\,dB attenuator and another 10\,dB attenuator at the output of the second LNA, which in fact remove all the extra added gain, was developed and tested in the laboratory and under the sky (but not deployed for longer data collection).
In this configuration the reflection coefficient S22 measured at the output of the front-end is almost exactly the same whether the input of the new front-end is terminated, opened or connected to the antenna (Fig.~\ref{fig_s22_morgan_fe}).
The initial tests of the new front-end in the chamber and under the sky showed that the $\approx$1.275\,MHz ripple and any other effects down from the output of the front-end can be calibrated out.
Further measurements and analysis will assess whether suppression of impedance mismatch propagation effects is already sufficient in this configuration.
}

\red{
Furthermore, a noise source was added to enable ``hot/cold load'' calibration schema and continuously determine instantaneous receiver noise temperature without relying on a laboratory measurement and the assumption that it remains constant over the measurement duration.
Additionally, the two-position lossy ($\sim$0.75~dB) solid-state RF-switch has been replaced with a very low loss mechanical RF-switch.
A solid state RF-switch was used in the prototype system mainly for practical reasons. At the time we wanted to power the front-end with a single 5\,V voltage (required by the LNA) whilst off-the-shelf mechanical RF-switches typically require at least 12\,V and draw much larger current.
The solid-state switch, besides its relatively high loss, is a more complex device (possibly not entirely passive) than a mechanical RF-switch.
Based on experiences with the prototype system we decided we should replace it by a simpler mechanical RF-switch in order to further simplify the signal path and remove undesired attenuation at the beginning of the signal path.
The above changes should lead to significant improvements in the calibration procedure and be more robust to the effects of the source impedance propagation effects.
}

\subsection{Conical log spiral antenna}
\label{subsec_antenna}


\begin{figure}
 \begin{center}
    \leavevmode
	 \includegraphics[width=3in,height=3.96in]{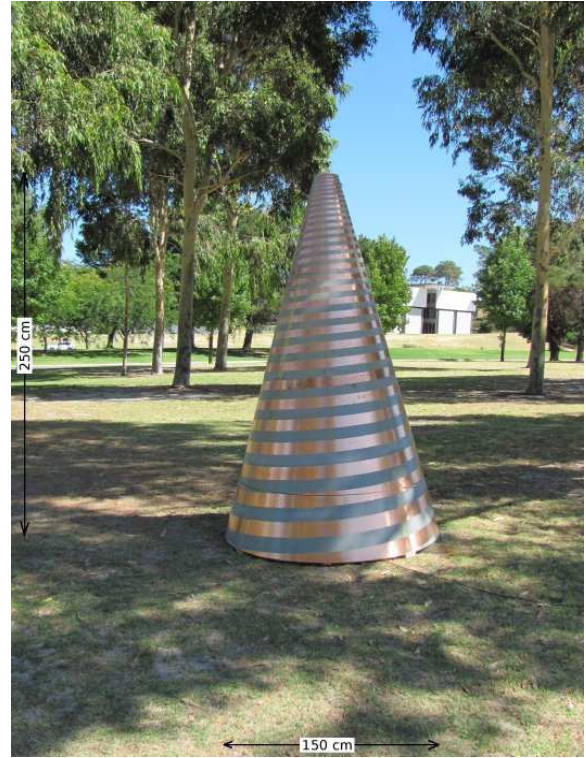}
    \caption{The conical log spiral antenna (before painting) during tests at Curtin University.}
    \label{fig_cone_notpainted}
  \end{center}
\end{figure}

\begin{figure}
 \begin{center}
    \leavevmode
	 \includegraphics[width=3in]{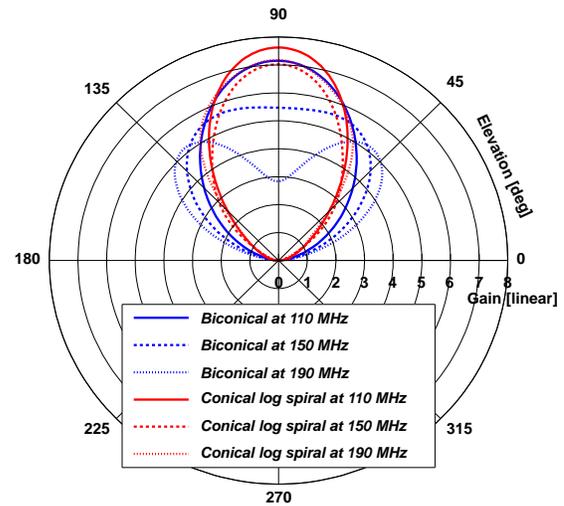}
    \caption{Comparison of simulated antenna pattern of the biconical and conical log spiral antennas at 110, 150 and 190 MHz.}
    \label{fig_pattern_cone_vs_bicon}
  \end{center}
\end{figure}

\red{
The off-the-shelf biconical antenna is going to be replaced with a bespoke conical log spiral antenna (Fig.~\ref{fig_cone_notpainted}), which is better matched over a wide frequency band.
Such an antenna will allow us to characterise all its components which should significantly improve the accuracy of the absolute calibration.
Moreover, it is a left-hand circularly polarised (LHCP) antenna which will help in suppressing reception of undesired right-hand circularly polarised (RHCP) signals from the ORBCOMM satellites.
With a narrower beam (FWHM $\approx70-80\degree$ at all frequencies) and low gain towards the horizon (Fig.~\ref{fig_pattern_cone_vs_bicon}), it will also enable us to use only data collected when the antenna is pointing towards the coldest regions of the sky.
}

\red{
Initial tests showed that the antenna is well matched (reflection coefficient $\sim$-20\,dB) in a wider frequency range $\sim70-180$\,MHz. Therefore, equipped with a better matched antenna and experience gained with a prototype system we should be able to parameterise the reflected noise without any undesired attenuation between the antenna and receiver.
Thus, we plan to minimise losses between the antenna and the first LNA due to cable, RF-switch and in particular remove the 3\,dB attenuator.
}

\section{Conclusions}
\label{discussion_and_future}



\red{
The precision required for the global EoR experiment is certainly a very challenging task from the instrumental and engineering points of view, for exceeding specifications of typical low radio frequency applications (including radio astronomy).
It requires unprecedented calibration precision and thus pushes boundaries of the radiometer techniques at low radio frequencies. 
The prototype BIGHORNS system allowed us to learn multiple lessons, which triggered modifications in the initial signal path and will inform future deployments.
The experiences from several deployments and laboratory tests once again confirmed the importance of a radio quiet locations for the global EoR experiment.
Nevertheless, even in superb radio quiet locations there are undesired effects such as tropospheric ducting or high power ORBCOMM communication satellites which significantly affect the data and require excision of a significant fraction of the data (at least 20\%). 
In the case of the data presented here the excision rate was $\sim$20\%, but it may become even higher once more rigorous criteria are applied to further suppress undesired effects of ORBCOMM satellites transmission. }

\red{Moreover, in the best parts of the year (when the Galactic Centre is in phase with the Sun) it is possible to collect maximally (without taking into account any excision due to RFI or other effects) about 5 hours of data in optimal conditions (elevation of the Galactic Centre $\le-20 \degree$) and thus accuracy $\sim$10\,mK can be achieved in less than two nights.
On the other hand in the worst parts of the year even 6 nights are required to integrate down to $\sim$10\,mK. The shorter integration time the better because it is much more likely to get two consecutive nights with similar conditions (weather, ionosphere etc.).
Due to the above, the system itself must be extremely stable and not introduce additional variability which is already present due to environmental and propagation (ionosphere) factors.}
\rednew{The stability of the existing system has been extensively tested and seems to be sufficient to integrate down the thermal noise to the desired precision (Figures \ref{fig_xmass_ref_rms_vs_time} and \ref{fig_delta_trec_cold_sky}).
Nevertheless, further sophisticated data analysis is required in order to separate the global EoR signal from the foregrounds (for example such as presented by \citet{2013PhRvD..87d3002L}).}

\red{
The self-generated emission from the system was measured in an anechoic chamber in order to ensure that detection of the global EoR will not be compromised by the system's own noise.
Although no particularly significant emissions have been identified, the results of the tests reassured us that these issues should be taken seriously in an experiment requiring such extreme precision.
In the case of the presented system the $\sim$100\,m separation between the antenna and the back-end electronics should be sufficient to suppress the EMI to a level not affecting the global EoR detection. 
However, if possible, any amount of extra shielding (terrain or shielded building) should be used to further suppress the noise generated in the back-end electronics.}

\red{Customised software has been developed to automatically reduce and calibrate the data, control their quality and select integrations which can be used for the global EoR data analysis. 
In order to test understanding of the sky signal and assess quality of the calibration, a procedure integrating the FEKO antenna model with the sky model was developed. 
The results of the absolute calibration procedure agree with the model predictions to within 10\%, which can be attributed to several factors such as incomplete knowledge of the off-the-shelf
biconical antenna characteristics and noise reflections between the antenna and the receiver, uncertainty of the sky model or inaccuracies in the calibration procedure.}

\red{Based on the collected experiences, several weak points in the existing signal path, which limit calibration accuracy, have been identified.
They are mainly due to ``downstream propagation effects'' related to the impedance of the source (antenna or 50\,$\Omega$ terminator) connected to the front-end changing its output properties (S22) and affecting standing noise waves downstream in the signal path such as ripple due to the 100\,m cable.
This ripple cannot be fully calibrated out and can only be modelled which significantly complicates the data analysis.
Initial tests have proven that the problem can be dealt with by adding more components (attenuators and amplifiers) to the front-end which more efficiently ``isolate'' its output from the input.
This modification together with several other improvements in the front-end receiver such as change of solid-state RF-switch to a mechanical RF-switch and 
addition of the noise source implementing ``hot/cold'' calibration schema should significantly improve accuracy of the calibration. 
The main reasons to replace the solid-state RF-switch with a mechanical RF-switch were to remove undesired attenuation at the very beginning of the signal path and to further simplify the signal path.
Generally our experiences with the prototype system indicate that simplicity of the signal path is essential in the global EoR experiment.
Even a simple system has many internal and external parameters which have to be controlled to high precision, and introduction of more complex elements leads to more complex effects in the system.
}

\red{
Finally, initial tests (such as reflection coefficient measurements) of the custom designed and built conical log spiral antenna developed specifically for this experiment are very promising.
It is expected to be well matched in the wider frequency band. Moreover, having a bespoke instrument provided more control over its components (such as balun, cables etc.).
Therefore, they can be accessed, changed or characterised easier than in the case of the off-the-shelf product.
}

\red{
The BIGHORNS system in its current form allows the collection of good quality data to probe the possibility of detecting the global EoR signal. The data collected in remote locations are being analysed.
We expect that the final system equipped with a conical log spiral antenna and an improved receiver will sufficiently improve calibration accuracy to constrain the cosmological models or even detect the Epoch of Reionisation.
}




\begin{acknowledgements}
This research was conducted by the Australian Research Council Centre of Excellence for All-sky Astrophysics (CAASTRO), through project number CE110001020.
The International Centre for Radio Astronomy Research (ICRAR) is a Joint Venture between Curtin University and the University of Western Australia, funded by the State Government of Western Australia and the Joint Venture partners.
Some of the results in this paper have been derived using the HEALPix \citep{2005ApJ...622..759G}.
We would like to thank Andr\'e Offringa for customising AOflagger software to work with BIGHORNS data.
\end{acknowledgements}


\bibliographystyle{apj}
\bibliography{refs}

\begin{thebibliography}{38}
\expandafter\ifx\csname natexlab\endcsname\relax\def\natexlab#1{#1}\fi

\bibitem[{{Bernardi} {et~al.}(2014){Bernardi}, {McQuinn}, \&
  {Greenhill}}]{2014arXiv1404.0887B}
{Bernardi}, G., {McQuinn}, M., \& {Greenhill}, L.~J. 2014, ArXiv e-prints

\bibitem[{{Bowman} \& {Rogers}(2010)}]{2010Natur.468..796B}
{Bowman}, J.~D. \& {Rogers}, A.~E.~E. 2010, \nat, 468, 796

\bibitem[{{Bray} {et~al.}(2013){Bray}, {Ekers}, \&
  {Roberts}}]{2013ExA....36..155B}
{Bray}, J.~D., {Ekers}, R.~D., \& {Roberts}, P. 2013, Experimental Astronomy,
  36, 155

\bibitem[{{Burns} {et~al.}(2012){Burns}, {Lazio}, {Bale}, {Bowman}, {Bradley},
  {Carilli}, {Furlanetto}, {Harker}, {Loeb}, \&
  {Pritchard}}]{2012AdSpR..49..433B}
{Burns}, J.~O., {Lazio}, J., {Bale}, S., {Bowman}, J., {Bradley}, R.,
  {Carilli}, C., {Furlanetto}, S., {Harker}, G., {Loeb}, A., \& {Pritchard}, J.
  2012, Advances in Space Research, 49, 433

\bibitem[{{Chippendale}(2009)}]{2009PhDT.......301C}
{Chippendale}, A.~P.~. 2009, PhD thesis, CSIRO Astronomy and Space Science

\bibitem[{CISPR 22(2008)}]{CISPR_STD}
CISPR 22. 2008, Information technology equipment - Radio disturbance
  characteristics - Limits and methods of measurement, 6th edn., International
  Electrotechnical Commission

\bibitem[{{Datta} {et~al.}(2014){Datta}, {Bradley}, {Burns}, {Harker},
  {Komjathy}, \& {Lazio}}]{2014arXiv1409.0513D}
{Datta}, A., {Bradley}, R., {Burns}, J.~O., {Harker}, G., {Komjathy}, A., \&
  {Lazio}, T.~J.~W. 2014, ArXiv e-prints

\bibitem[{{de Oliveira-Costa} {et~al.}(2008){de Oliveira-Costa}, {Tegmark},
  {Gaensler}, {Jonas}, {Landecker}, \& {Reich}}]{2008MNRAS.388..247D}
{de Oliveira-Costa}, A., {Tegmark}, M., {Gaensler}, B.~M., {Jonas}, J.,
  {Landecker}, T.~L., \& {Reich}, P. 2008, \mnras, 388, 247

\bibitem[{{Furlanetto} {et~al.}(2006){Furlanetto}, {Oh}, \&
  {Briggs}}]{2006PhR...433..181F}
{Furlanetto}, S.~R., {Oh}, S.~P., \& {Briggs}, F.~H. 2006, \physrep, 433, 181

\bibitem[{{G{\'o}rski} {et~al.}(2005){G{\'o}rski}, {Hivon}, {Banday},
  {Wandelt}, {Hansen}, {Reinecke}, \& {Bartelmann}}]{2005ApJ...622..759G}
{G{\'o}rski}, K.~M., {Hivon}, E., {Banday}, A.~J., {Wandelt}, B.~D., {Hansen},
  F.~K., {Reinecke}, M., \& {Bartelmann}, M. 2005, \apj, 622, 759

\bibitem[{{Greenhill} {et~al.}(2014){Greenhill}, {Kocz}, {Barsdell}, {Clark},
  \& {LEDA Collaboration}}]{2014era..conf10301G}
{Greenhill}, L.~J., {Kocz}, J., {Barsdell}, B.~R., {Clark}, M.~A., \& {LEDA
  Collaboration}. 2014, in Exascale Radio Astronomy, 10301

\bibitem[{{Harker} {et~al.}(2012){Harker}, {Pritchard}, {Burns}, \&
  {Bowman}}]{2012MNRAS.419.1070H}
{Harker}, G.~J.~A., {Pritchard}, J.~R., {Burns}, J.~O., \& {Bowman}, J.~D.
  2012, \mnras, 419, 1070

\bibitem[{Hitney {et~al.}(1985)Hitney, Richter, Pappert, Anderson, \&
  Baumgartner}]{1457406}
Hitney, H., Richter, J., Pappert, R., Anderson, K., \& Baumgartner, G.B., J.
  1985, Proceedings of the IEEE, 73, 265

\bibitem[{ITU-R P.526-12(2012)}]{ITUR_MODEL}
ITU-R P.526-12. 2012, Propagation by diffraction, International
  Telecommunication Union Radiocommunication Sector

\bibitem[{{James} {et~al.}(2011){James}, {Protheroe}, {Ekers},
  {Alvarez-Mu{\~n}iz}, {McFadden}, {Phillips}, {Roberts}, \&
  {Bray}}]{2011MNRAS.410..885J}
{James}, C.~W., {Protheroe}, R.~J., {Ekers}, R.~D., {Alvarez-Mu{\~n}iz}, J.,
  {McFadden}, R.~A., {Phillips}, C.~J., {Roberts}, P., \& {Bray}, J.~D. 2011,
  \mnras, 410, 885

\bibitem[{{Johnston} \& {McRory}(1998)}]{wheeler_cap}
{Johnston}, R.~H. \& {McRory}, J. 1998, Antennas and Propagation Magazine,
  IEEE, 40

\bibitem[{{Liu} {et~al.}(2013){Liu}, {Pritchard}, {Tegmark}, \&
  {Loeb}}]{2013PhRvD..87d3002L}
{Liu}, A., {Pritchard}, J.~R., {Tegmark}, M., \& {Loeb}, A. 2013, \prd, 87,
  043002

\bibitem[{{Mellema} {et~al.}(2013){Mellema}, {Koopmans}, {Abdalla}, {Bernardi},
  {Ciardi}, {Daiboo}, {de Bruyn}, {Datta}, {Falcke}, {Ferrara}, {Iliev},
  {Iocco}, {Jeli{\'c}}, {Jensen}, {Joseph}, {Labroupoulos}, {Meiksin},
  {Mesinger}, {Offringa}, {Pandey}, {Pritchard}, {Santos}, {Schwarz},
  {Semelin}, {Vedantham}, {Yatawatta}, \& {Zaroubi}}]{2013ExA....36..235M}
{Mellema}, G., {Koopmans}, L.~V.~E., {Abdalla}, F.~A., {Bernardi}, G.,
  {Ciardi}, B., {Daiboo}, S., {de Bruyn}, A.~G., {Datta}, K.~K., {Falcke}, H.,
  {Ferrara}, A., {Iliev}, I.~T., {Iocco}, F., {Jeli{\'c}}, V., {Jensen}, H.,
  {Joseph}, R., {Labroupoulos}, P., {Meiksin}, A., {Mesinger}, A., {Offringa},
  A.~R., {Pandey}, V.~N., {Pritchard}, J.~R., {Santos}, M.~G., {Schwarz},
  D.~J., {Semelin}, B., {Vedantham}, H., {Yatawatta}, S., \& {Zaroubi}, S.
  2013, Experimental Astronomy, 36, 235

\bibitem[{MIL-STD461F(2007)}]{MILSTD}
MIL-STD461F. 2007, Requirements for the Control of Electromagnetic Interference
  Characteristics of Subsystems and Equipment, Department of Defense

\bibitem[{{Mirocha} {et~al.}(2013){Mirocha}, {Harker}, \&
  {Burns}}]{2013ApJ...777..118M}
{Mirocha}, J., {Harker}, G.~J.~A., \& {Burns}, J.~O. 2013, \apj, 777, 118

\bibitem[{{Morales} \& {Wyithe}(2010)}]{2010ARA&A..48..127M}
{Morales}, M.~F. \& {Wyithe}, J.~S.~B. 2010, \araa, 48, 127

\bibitem[{{Morandi} \& {Barkana}(2012)}]{2012MNRAS.424.2551M}
{Morandi}, A. \& {Barkana}, R. 2012, \mnras, 424, 2551

\bibitem[{{Offringa} {et~al.}(2012){Offringa}, {van de Gronde}, \&
  {Roerdink}}]{2012A&A...539A..95O}
{Offringa}, A.~R., {van de Gronde}, J.~J., \& {Roerdink}, J.~B.~T.~M. 2012,
  \aap, 539, A95

\bibitem[{{Parsons} {et~al.}(2010){Parsons}, {Backer}, {Foster}, {Wright},
  {Bradley}, {Gugliucci}, {Parashare}, {Benoit}, {Aguirre}, {Jacobs},
  {Carilli}, {Herne}, {Lynch}, {Manley}, \& {Werthimer}}]{2010AJ....139.1468P}
{Parsons}, A.~R., {Backer}, D.~C., {Foster}, G.~S., {Wright}, M.~C.~H.,
  {Bradley}, R.~F., {Gugliucci}, N.~E., {Parashare}, C.~R., {Benoit}, E.~E.,
  {Aguirre}, J.~E., {Jacobs}, D.~C., {Carilli}, C.~L., {Herne}, D., {Lynch},
  M.~J., {Manley}, J.~R., \& {Werthimer}, D.~J. 2010, \aj, 139, 1468

\bibitem[{{Patra} {et~al.}(2013){Patra}, {Subrahmanyan}, {Raghunathan}, \&
  {Udaya Shankar}}]{2013ExA....36..319P}
{Patra}, N., {Subrahmanyan}, R., {Raghunathan}, A., \& {Udaya Shankar}, N.
  2013, Experimental Astronomy, 36, 319

\bibitem[{{Pence} {et~al.}(2010){Pence}, {Chiappetti}, {Page}, {Shaw}, \&
  {Stobie}}]{2010A&A...524A..42P}
{Pence}, W.~D., {Chiappetti}, L., {Page}, C.~G., {Shaw}, R.~A., \& {Stobie}, E.
  2010, \aap, 524, A42

\bibitem[{{Pritchard} \& {Loeb}(2008)}]{2008PhRvD..78j3511P}
{Pritchard}, J.~R. \& {Loeb}, A. 2008, \prd, 78, 103511

\bibitem[{{Pritchard} \& {Loeb}(2010)}]{2010PhRvD..82b3006P}
---. 2010, \prd, 82, 023006

\bibitem[{{Pritchard} \& {Loeb}(2012)}]{2012RPPh...75h6901P}
---. 2012, Reports on Progress in Physics, 75, 086901

\bibitem[{{Rogers} \& {Bowman}(2012)}]{2012RaSc...47.0K06R}
{Rogers}, A.~E.~E. \& {Bowman}, J.~D. 2012, Radio Science, 47, 0

\bibitem[{{Shaver} {et~al.}(1999){Shaver}, {Windhorst}, {Madau}, \& {de
  Bruyn}}]{1999A&A...345..380S}
{Shaver}, P.~A., {Windhorst}, R.~A., {Madau}, P., \& {de Bruyn}, A.~G. 1999,
  \aap, 345, 380

\bibitem[{{Taylor} {et~al.}(2012){Taylor}, {Ellingson}, {Kassim}, {Craig},
  {Dowell}, {Wolfe}, {Hartman}, {Bernardi}, {Clarke}, {Cohen}, {Dalal},
  {Erickson}, {Hicks}, {Greenhill}, {Jacoby}, {Lane}, {Lazio}, {Mitchell},
  {Navarro}, {Ord}, {Pihlstr{\"o}m}, {Polisensky}, {Ray}, {Rickard},
  {Schinzel}, {Schmitt}, {Sigman}, {Soriano}, {Stewart}, {Stovall}, {Tremblay},
  {Wang}, {Weiler}, {White}, \& {Wood}}]{2012JAI.....150004T}
{Taylor}, G.~B., {Ellingson}, S.~W., {Kassim}, N.~E., {Craig}, J., {Dowell},
  J., {Wolfe}, C.~N., {Hartman}, J., {Bernardi}, G., {Clarke}, T., {Cohen}, A.,
  {Dalal}, N.~P., {Erickson}, W.~C., {Hicks}, B., {Greenhill}, L.~J., {Jacoby},
  B., {Lane}, W., {Lazio}, J., {Mitchell}, D., {Navarro}, R., {Ord}, S.~M.,
  {Pihlstr{\"o}m}, Y., {Polisensky}, E., {Ray}, P.~S., {Rickard}, L.~J.,
  {Schinzel}, F.~K., {Schmitt}, H., {Sigman}, E., {Soriano}, M., {Stewart},
  K.~P., {Stovall}, K., {Tremblay}, S., {Wang}, D., {Weiler}, K.~W., {White},
  S., \& {Wood}, D.~L. 2012, Journal of Astronomical Instrumentation, 1, 50004

\bibitem[{{Tingay} {et~al.}(2013){Tingay}, {Goeke}, {Bowman}, {Emrich}, {Ord},
  {Mitchell}, {Morales}, {Booler}, {Crosse}, {Wayth}, {Lonsdale}, {Tremblay},
  {Pallot}, {Colegate}, {Wicenec}, {Kudryavtseva}, {Arcus}, {Barnes},
  {Bernardi}, {Briggs}, {Burns}, {Bunton}, {Cappallo}, {Corey}, {Deshpande},
  {Desouza}, {Gaensler}, {Greenhill}, {Hall}, {Hazelton}, {Herne}, {Hewitt},
  {Johnston-Hollitt}, {Kaplan}, {Kasper}, {Kincaid}, {Koenig}, {Kratzenberg},
  {Lynch}, {Mckinley}, {Mcwhirter}, {Morgan}, {Oberoi}, {Pathikulangara},
  {Prabu}, {Remillard}, {Rogers}, {Roshi}, {Salah}, {Sault}, {Udaya-Shankar},
  {Schlagenhaufer}, {Srivani}, {Stevens}, {Subrahmanyan}, {Waterson},
  {Webster}, {Whitney}, {Williams}, {Williams}, \&
  {Wyithe}}]{2013PASA...30....7T}
{Tingay}, S.~J., {Goeke}, R., {Bowman}, J.~D., {Emrich}, D., {Ord}, S.~M.,
  {Mitchell}, D.~A., {Morales}, M.~F., {Booler}, T., {Crosse}, B., {Wayth},
  R.~B., {Lonsdale}, C.~J., {Tremblay}, S., {Pallot}, D., {Colegate}, T.,
  {Wicenec}, A., {Kudryavtseva}, N., {Arcus}, W., {Barnes}, D., {Bernardi}, G.,
  {Briggs}, F., {Burns}, S., {Bunton}, J.~D., {Cappallo}, R.~J., {Corey},
  B.~E., {Deshpande}, A., {Desouza}, L., {Gaensler}, B.~M., {Greenhill}, L.~J.,
  {Hall}, P.~J., {Hazelton}, B.~J., {Herne}, D., {Hewitt}, J.~N.,
  {Johnston-Hollitt}, M., {Kaplan}, D.~L., {Kasper}, J.~C., {Kincaid}, B.~B.,
  {Koenig}, R., {Kratzenberg}, E., {Lynch}, M.~J., {Mckinley}, B., {Mcwhirter},
  S.~R., {Morgan}, E., {Oberoi}, D., {Pathikulangara}, J., {Prabu}, T.,
  {Remillard}, R.~A., {Rogers}, A.~E.~E., {Roshi}, A., {Salah}, J.~E., {Sault},
  R.~J., {Udaya-Shankar}, N., {Schlagenhaufer}, F., {Srivani}, K.~S.,
  {Stevens}, J., {Subrahmanyan}, R., {Waterson}, M., {Webster}, R.~L.,
  {Whitney}, A.~R., {Williams}, A., {Williams}, C.~L., \& {Wyithe}, J.~S.~B.
  2013, \pasa, 30, 7

\bibitem[{{van Haarlem} {et~al.}(2013){van Haarlem}, {Wise}, {Gunst}, {Heald},
  {McKean}, {Hessels}, {de Bruyn}, {Nijboer}, {Swinbank}, {Fallows},
  {Brentjens}, {Nelles}, {Beck}, {Falcke}, {Fender}, {H{\"o}randel},
  {Koopmans}, {Mann}, {Miley}, {R{\"o}ttgering}, {Stappers}, {Wijers},
  {Zaroubi}, {van den Akker}, {Alexov}, {Anderson}, {Anderson}, {van Ardenne},
  {Arts}, {Asgekar}, {Avruch}, {Batejat}, {B{\"a}hren}, {Bell}, {Bell}, {van
  Bemmel}, {Bennema}, {Bentum}, {Bernardi}, {Best}, {B{\^i}rzan}, {Bonafede},
  {Boonstra}, {Braun}, {Bregman}, {Breitling}, {van de Brink}, {Broderick},
  {Broekema}, {Brouw}, {Br{\"u}ggen}, {Butcher}, {van Cappellen}, {Ciardi},
  {Coenen}, {Conway}, {Coolen}, {Corstanje}, {Damstra}, {Davies}, {Deller},
  {Dettmar}, {van Diepen}, {Dijkstra}, {Donker}, {Doorduin}, {Dromer}, {Drost},
  {van Duin}, {Eisl{\"o}ffel}, {van Enst}, {Ferrari}, {Frieswijk}, {Gankema},
  {Garrett}, {de Gasperin}, {Gerbers}, {de Geus}, {Grie{\ss}meier}, {Grit},
  {Gruppen}, {Hamaker}, {Hassall}, {Hoeft}, {Holties}, {Horneffer}, {van der
  Horst}, {van Houwelingen}, {Huijgen}, {Iacobelli}, {Intema}, {Jackson},
  {Jelic}, {de Jong}, {Juette}, {Kant}, {Karastergiou}, {Koers}, {Kollen},
  {Kondratiev}, {Kooistra}, {Koopman}, {Koster}, {Kuniyoshi}, {Kramer},
  {Kuper}, {Lambropoulos}, {Law}, {van Leeuwen}, {Lemaitre}, {Loose}, {Maat},
  {Macario}, {Markoff}, {Masters}, {McFadden}, {McKay-Bukowski}, {Meijering},
  {Meulman}, {Mevius}, {Middelberg}, {Millenaar}, {Miller-Jones}, {Mohan},
  {Mol}, {Morawietz}, {Morganti}, {Mulcahy}, {Mulder}, {Munk}, {Nieuwenhuis},
  {van Nieuwpoort}, {Noordam}, {Norden}, {Noutsos}, {Offringa}, {Olofsson},
  {Omar}, {Orr{\'u}}, {Overeem}, {Paas}, {Pandey-Pommier}, {Pandey}, {Pizzo},
  {Polatidis}, {Rafferty}, {Rawlings}, {Reich}, {de Reijer}, {Reitsma},
  {Renting}, {Riemers}, {Rol}, {Romein}, {Roosjen}, {Ruiter}, {Scaife}, {van
  der Schaaf}, {Scheers}, {Schellart}, {Schoenmakers}, {Schoonderbeek},
  {Serylak}, {Shulevski}, {Sluman}, {Smirnov}, {Sobey}, {Spreeuw}, {Steinmetz},
  {Sterks}, {Stiepel}, {Stuurwold}, {Tagger}, {Tang}, {Tasse}, {Thomas},
  {Thoudam}, {Toribio}, {van der Tol}, {Usov}, {van Veelen}, {van der Veen},
  {ter Veen}, {Verbiest}, {Vermeulen}, {Vermaas}, {Vocks}, {Vogt}, {de Vos},
  {van der Wal}, {van Weeren}, {Weggemans}, {Weltevrede}, {White}, {Wijnholds},
  {Wilhelmsson}, {Wucknitz}, {Yatawatta}, {Zarka}, {Zensus}, \& {van
  Zwieten}}]{2013A&A...556A...2V}
{van Haarlem}, M.~P., {Wise}, M.~W., {Gunst}, A.~W., {Heald}, G., {McKean},
  J.~P., {Hessels}, J.~W.~T., {de Bruyn}, A.~G., {Nijboer}, R., {Swinbank}, J.,
  {Fallows}, R., {Brentjens}, M., {Nelles}, A., {Beck}, R., {Falcke}, H.,
  {Fender}, R., {H{\"o}randel}, J., {Koopmans}, L.~V.~E., {Mann}, G., {Miley},
  G., {R{\"o}ttgering}, H., {Stappers}, B.~W., {Wijers}, R.~A.~M.~J.,
  {Zaroubi}, S., {van den Akker}, M., {Alexov}, A., {Anderson}, J., {Anderson},
  K., {van Ardenne}, A., {Arts}, M., {Asgekar}, A., {Avruch}, I.~M., {Batejat},
  F., {B{\"a}hren}, L., {Bell}, M.~E., {Bell}, M.~R., {van Bemmel}, I.,
  {Bennema}, P., {Bentum}, M.~J., {Bernardi}, G., {Best}, P., {B{\^i}rzan}, L.,
  {Bonafede}, A., {Boonstra}, A.-J., {Braun}, R., {Bregman}, J., {Breitling},
  F., {van de Brink}, R.~H., {Broderick}, J., {Broekema}, P.~C., {Brouw},
  W.~N., {Br{\"u}ggen}, M., {Butcher}, H.~R., {van Cappellen}, W., {Ciardi},
  B., {Coenen}, T., {Conway}, J., {Coolen}, A., {Corstanje}, A., {Damstra}, S.,
  {Davies}, O., {Deller}, A.~T., {Dettmar}, R.-J., {van Diepen}, G.,
  {Dijkstra}, K., {Donker}, P., {Doorduin}, A., {Dromer}, J., {Drost}, M., {van
  Duin}, A., {Eisl{\"o}ffel}, J., {van Enst}, J., {Ferrari}, C., {Frieswijk},
  W., {Gankema}, H., {Garrett}, M.~A., {de Gasperin}, F., {Gerbers}, M., {de
  Geus}, E., {Grie{\ss}meier}, J.-M., {Grit}, T., {Gruppen}, P., {Hamaker},
  J.~P., {Hassall}, T., {Hoeft}, M., {Holties}, H.~A., {Horneffer}, A., {van
  der Horst}, A., {van Houwelingen}, A., {Huijgen}, A., {Iacobelli}, M.,
  {Intema}, H., {Jackson}, N., {Jelic}, V., {de Jong}, A., {Juette}, E.,
  {Kant}, D., {Karastergiou}, A., {Koers}, A., {Kollen}, H., {Kondratiev},
  V.~I., {Kooistra}, E., {Koopman}, Y., {Koster}, A., {Kuniyoshi}, M.,
  {Kramer}, M., {Kuper}, G., {Lambropoulos}, P., {Law}, C., {van Leeuwen}, J.,
  {Lemaitre}, J., {Loose}, M., {Maat}, P., {Macario}, G., {Markoff}, S.,
  {Masters}, J., {McFadden}, R.~A., {McKay-Bukowski}, D., {Meijering}, H.,
  {Meulman}, H., {Mevius}, M., {Middelberg}, E., {Millenaar}, R.,
  {Miller-Jones}, J.~C.~A., {Mohan}, R.~N., {Mol}, J.~D., {Morawietz}, J.,
  {Morganti}, R., {Mulcahy}, D.~D., {Mulder}, E., {Munk}, H., {Nieuwenhuis},
  L., {van Nieuwpoort}, R., {Noordam}, J.~E., {Norden}, M., {Noutsos}, A.,
  {Offringa}, A.~R., {Olofsson}, H., {Omar}, A., {Orr{\'u}}, E., {Overeem}, R.,
  {Paas}, H., {Pandey-Pommier}, M., {Pandey}, V.~N., {Pizzo}, R., {Polatidis},
  A., {Rafferty}, D., {Rawlings}, S., {Reich}, W., {de Reijer}, J.-P.,
  {Reitsma}, J., {Renting}, G.~A., {Riemers}, P., {Rol}, E., {Romein}, J.~W.,
  {Roosjen}, J., {Ruiter}, M., {Scaife}, A., {van der Schaaf}, K., {Scheers},
  B., {Schellart}, P., {Schoenmakers}, A., {Schoonderbeek}, G., {Serylak}, M.,
  {Shulevski}, A., {Sluman}, J., {Smirnov}, O., {Sobey}, C., {Spreeuw}, H.,
  {Steinmetz}, M., {Sterks}, C.~G.~M., {Stiepel}, H.-J., {Stuurwold}, K.,
  {Tagger}, M., {Tang}, Y., {Tasse}, C., {Thomas}, I., {Thoudam}, S.,
  {Toribio}, M.~C., {van der Tol}, B., {Usov}, O., {van Veelen}, M., {van der
  Veen}, A.-J., {ter Veen}, S., {Verbiest}, J.~P.~W., {Vermeulen}, R.,
  {Vermaas}, N., {Vocks}, C., {Vogt}, C., {de Vos}, M., {van der Wal}, E., {van
  Weeren}, R., {Weggemans}, H., {Weltevrede}, P., {White}, S., {Wijnholds},
  S.~J., {Wilhelmsson}, T., {Wucknitz}, O., {Yatawatta}, S., {Zarka}, P.,
  {Zensus}, A., \& {van Zwieten}, J. 2013, \aap, 556, A2

\bibitem[{{Vedantham} {et~al.}(2014){Vedantham}, {Koopmans}, {de Bruyn},
  {Wijnholds}, {Ciardi}, \& {Brentjens}}]{2014MNRAS.437.1056V}
{Vedantham}, H.~K., {Koopmans}, L.~V.~E., {de Bruyn}, A.~G., {Wijnholds},
  S.~J., {Ciardi}, B., \& {Brentjens}, M.~A. 2014, \mnras, 437, 1056

\bibitem[{{Voytek} {et~al.}(2014){Voytek}, {Natarajan}, {J{\'a}uregui
  Garc{\'{\i}}a}, {Peterson}, \& {L{\'o}pez-Cruz}}]{2014ApJ...782L...9V}
{Voytek}, T.~C., {Natarajan}, A., {J{\'a}uregui Garc{\'{\i}}a}, J.~M.,
  {Peterson}, J.~B., \& {L{\'o}pez-Cruz}, O. 2014, \apjl, 782, L9

\bibitem[{Wyithe(2014)}]{AASKA14}
Wyithe, S. 2014, in Advancing Astrophysics with the Square Kilometre Array,
  Proceedings of Science

\bibitem[{{Zahn} {et~al.}(2012){Zahn}, {Reichardt}, {Shaw}, {Lidz}, {Aird},
  {Benson}, {Bleem}, {Carlstrom}, {Chang}, {Cho}, {Crawford}, {Crites}, {de
  Haan}, {Dobbs}, {Dor{\'e}}, {Dudley}, {George}, {Halverson}, {Holder},
  {Holzapfel}, {Hoover}, {Hou}, {Hrubes}, {Joy}, {Keisler}, {Knox}, {Lee},
  {Leitch}, {Lueker}, {Luong-Van}, {McMahon}, {Mehl}, {Meyer}, {Millea},
  {Mohr}, {Montroy}, {Natoli}, {Padin}, {Plagge}, {Pryke}, {Ruhl}, {Schaffer},
  {Shirokoff}, {Spieler}, {Staniszewski}, {Stark}, {Story}, {van Engelen},
  {Vanderlinde}, {Vieira}, \& {Williamson}}]{2012ApJ...756...65Z}
{Zahn}, O., {Reichardt}, C.~L., {Shaw}, L., {Lidz}, A., {Aird}, K.~A.,
  {Benson}, B.~A., {Bleem}, L.~E., {Carlstrom}, J.~E., {Chang}, C.~L., {Cho},
  H.~M., {Crawford}, T.~M., {Crites}, A.~T., {de Haan}, T., {Dobbs}, M.~A.,
  {Dor{\'e}}, O., {Dudley}, J., {George}, E.~M., {Halverson}, N.~W., {Holder},
  G.~P., {Holzapfel}, W.~L., {Hoover}, S., {Hou}, Z., {Hrubes}, J.~D., {Joy},
  M., {Keisler}, R., {Knox}, L., {Lee}, A.~T., {Leitch}, E.~M., {Lueker}, M.,
  {Luong-Van}, D., {McMahon}, J.~J., {Mehl}, J., {Meyer}, S.~S., {Millea}, M.,
  {Mohr}, J.~J., {Montroy}, T.~E., {Natoli}, T., {Padin}, S., {Plagge}, T.,
  {Pryke}, C., {Ruhl}, J.~E., {Schaffer}, K.~K., {Shirokoff}, E., {Spieler},
  H.~G., {Staniszewski}, Z., {Stark}, A.~A., {Story}, K., {van Engelen}, A.,
  {Vanderlinde}, K., {Vieira}, J.~D., \& {Williamson}, R. 2012, \apj, 756, 65

\end{thebibliography}

\end{document}